# Ultrafast symmetry modulation and induced magnetic excitation in the Kagome metal RbV$_3$Sb$_5$

Mengxue Guan[1,2†*], Xiaodong Zhou[3†], Jingyi Duan[1], Chaoxi Cui[1], Wei Jiang[1,2], Zeying Zhang[4], Binhua Zhang[5], Zhengwei Nie[6], Xun Shi[1,2], Zhiwei Wang[1,2], Yugui Yao[1,2*]

[1]Centre for Quantum Physics, Key Laboratory of Advanced Optoelectronic Quantum Architecture and Measurement (Ministry of Education), School of Physics, Beijing Institute of Technology, Beijing 100081, China

[2]International Center for Quantum Materials, Beijing Institute of Technology, Zhuhai, 519000, China

[3]School of Physical Science and Technology, Tiangong University, Tianjin 300387, China

[4]College of Mathematics and Physics, Beijing University of Chemical Technology, Beijing 100029, China

[5]Key Laboratory of Computational Physical Sciences (Ministry of Education), Institute of Computational Physical Sciences, State Key Laboratory of Surface Physics, and Department of Physics, Fudan University, Shanghai 200433, China

[6]Beijing National Laboratory for Condensed Matter Physics and Institute of Physics, Chinese Academy of Sciences, Beijing 100190, China

†These authors contributed equally: Mengxue Guan, Xiaodong Zhou

*Corresponding author: ygyao@bit.edu.cn, mxguan@bit.edu.cn

Light-matter interaction in frustrated Kagome metals enables access to hidden quantum states, yet the microscopic origin of symmetry breaking under ultrafast excitation remains elusive. Here, we uncover a microscopic mechanism for laser-induced symmetry breaking in RbV$_3$Sb$_5$ through first-principles real-time simulations. Selective excitation of a single-$Q_M$ phonon mode dynamically breaks both rotational and time-reversal symmetries within the 2 × 2× 1 charge density wave (CDW) superlattice. The resulting anisotropic lattice distortion lifts geometric frustration and stabilizes a nonequilibrium ferrimagnetic phase, accompanied by a sizable intrinsic anomalous Hall effect. Distinct from prior interpretations based on orbital antiferromagnetism or extrinsic perturbations, our findings reveal a spin-driven pathway for symmetry breaking under strong optical fields. These results provide a microscopic foundation for exploring how spin, lattice and charge degrees of freedom are intertwined in nonequilibrium correlated states.

*Introduction*

The Kagome lattice, composed of corner-sharing triangles, offers an ideal platform to explore the interplay between magnetism, topology, and strong electronic correlations[1]. Symmetry breaking plays a central role in tuning many-body interactions and facilitating the emergence of novel quantum phases. For instance, the antiferromagnetic Kagome spin model exhibits strong quantum fluctuations, rendering it a canonical example of magnetic frustration[2,3], with a highly degenerate ground-states analogous to liquid-ice configurations. This degeneracy can be lifted by introducing anisotropies[4,5], long-range interactions[6], structural distortions[7] or spin-lattice coupling[8], giving rise to distinct magnetic excitations[9,10]. Hence, precise control over the coupling between magnetic, charge and lattice degrees of freedom is essential.

$A$V$_3$Sb$_5$ ($A$ = K, Cs, Rb), a prototypical Kagome metal, hosts a rich array of phenomena including $Z_2$ topology, superconductivity, and charge density wave (CDW) order[11-13]. Above the CDW transition temperature ($T_{CDW} \approx 78 \sim 104$ K), $A$V$_3$Sb$_5$ is Pauli paramagnetic, yet multiple experiments, including muon spin relaxation ($\mu$SR) [14-16], magneto-transport[17-20] and scanning tunneling microscopy (STM) studies[21-23] indicate time-reversal symmetry breaking ($\mathcal{T}$-breaking) within the CDW phase. Fourier-transformed STM images reveal unequal intensities at the three inequivalent $M$-point superlattice peaks, and their relative ordering can be switched by an external magnetic field, consistent with a chiral CDW state [21-25]. In the absence of long-range magnetic order, this $\mathcal{T}$-breaking has been attributed to orbital antiferromagnetism, an imaginary CDW component stabilized under appropriate band filling[26-28], which distinguishes $A$V$_3$Sb$_5$ from conventional Kagome magnets[29,30]. Upon further cooling, the CDW phase also develops rotational-symmetry breaking, strengthening its coupling to other electronic orders and profoundly effects the superconducting properties[31-34].

Despite numerous reports of rotational-symmetry breaking and $\mathcal{T}$-breaking, their reproducibility and microscopic origin remain contested. The observed $\mathcal{C}_6$ to $\mathcal{C}_2$ anisotropy has been ascribed either to interlayer phase shifts[35,36] or to a bulk nematic transitions[31,32], both of which yield unequal STM Bragg-peak intensities; yet a recent study finds no spontaneous symmetry breaking in the absence of external strain or magnetic fields[37]. Likewise, the mechanism of $\mathcal{T}$-breaking, particularly the proposed orbital antiferromagnetism scenario, is yet to reach consensus[38-40].

In this work, we reveal tunable symmetry breaking rooted in selective phonon excitation within a well-defined 2 × 2 lattice, a universally acknowledged precursor to correlated states in $A$V$_3$Sb$_5$, and therefore, the above uncertainties can be circumvented. Enhancing a specific single-$Q_M$ phonon mode in the 2 × 2× 1 CDW superlattice of $A$V$_3$Sb$_5$ drives simultaneous breaking of threefold rotational and time-reversal symmetries. The resulting anisotropic electronic hopping among the three sublattices relieves geometric frustration and stabilizes a nonequilibrium ferrimagnetic state. The induced net spin moment, in conjunction with spin-orbit coupling, generates finite Berry curvature in momentum space, giving rise to an intrinsic anomalous Hall effect (AHE). Simulated STM spectra based on this model reproduce key features observed in pervious experiments[41], providing a promising foundation for field-tunable symmetry breaking and emergent quantum phases in Kagome lattices systems.

***Structure and symmetry of $AV_3Sb_5$:***

In its high-temperature pristine phase, $A$V$_3$Sb$_5$ crystallizes in a layered structure comprising V-Sb sheets intercalated by alkali metal atoms, with space group *P*6/mmm (little co-group *D$_{2h}$*) [Fig. 1(a)]. Phonon dispersion calculation reveals imaginary modes at the *M* and *L* points of the Brillouin zone [Fig. S2(a)]. The degenerate imaginary modes at the three symmetry-equivalent *M* points transform according to the irreducible representation $M_1^+$ [42]. The associated CDW order parameters, denoted $M_\alpha$ ($\alpha = 1,2,3$), correspond to nesting vectors $\boldsymbol{Q}_M^\alpha$ [Fig. 1(b-c)] and are primarily characterized by in-plane displacements of V atoms. A distortion localized to a single *M* point defines a single-$Q_M$ mode. When all three $M_\alpha$ share equal amplitude, a triple-$Q_M$ superposition forms, resulting in "Star of David" (SD) and/or "inverse SD" (ISD) patterns with a 2×2×1 superlattice[36]. These distortions preserve the same space group symmetry of the pristine structure, with the ISD configuration being energetically favored[43] [Fig. 1(d)].

The primary difference between the *M*- and *L*-point modes lies in their interlayer phase relation: $M_\alpha$ modes correspond to in-phase displacement along the *c*-axis, while $L_\alpha$ modes are out-of-phase. The coexistence of both sets of nesting vectors enables diverse stacking and interlayer modulation patterns[32,44-46]. Consequently, the three-dimensional CDW structures vary across $A$V$_3$Sb$_5$ compounds, and the resultant electronic phases are sensitive to temperature, doping, pressure, strain and surface effects, rendering the precise nature of the CDW phase a subject of ongoing debate[1,30,34]. In this work, we focus on the nonequilibrium dynamics of the 2 × 2 × 1 ISD phase, which serves as a fundamental CDW prototype[47], before extending to more complex stacking geometries.

### *Tuning rotational symmetry via laser excitation:*

Recent ultrafast laser experiments have revealed photoinduced lattice distortion in $A$V$_3$Sb$_5$, including selective breaking of the mirror plane perpendicular to the Kagome plane[41,46,48-50]. However, previous studies have primarily focused on symmetry-based phenomenology, leaving the microscopic mechanism behind these symmetry changes poorly understood. To explore the ultrafast response of $A$V$_3$Sb$_5$ to optical excitation, we employ real-time time-dependent density functional theory molecular dynamics (rt-TDDFT-MD) simulations. Taking RbV$_3$Sb$_5$ as a representative case, we apply 1025 nm laser pulses polarized along the in-plane V-V nearest-neighbor directions (i.e., along $\boldsymbol{Q}_M^\alpha$ in reciprocal space), to explore the origin of the observed photoinduced symmetry breaking[41]. Further computational details can be found in Note S1 of the Supplemental Material [51] and Refs. [52-68] therein.

The structural evolution obtained from rt-TDDFT simulations closely reproduces key experimental signatures. At the Brillouin zone center, the dominant Raman-active phonon involves out-of-plane oscillations of Sb2 atoms[49] [Fig. S3]. Additionally, in-plane breathing motions of V and Sb2 ions associated with CDW distortion, contribute significantly. In thermal equilibrium, these phonons belong to the fully symmetric $A_{1g}$ representation and couple with the amplitude modes of CDW phase [46,47]. However, upon laser excitation, anisotropic distortions of the V-V bond lengths emerge along the three single-$Q_M$ directions, depending sensitively on the laser polarization [Fig. S4].

Figure 2(a) presents the time evolution of the three single-$\mathbf{Q}_M$ mode amplitudes $M_\alpha (\alpha = 1,2,3)$ under excitation polarized along $\boldsymbol{Q}_M^1$ (i.e., the real space $a$-axis). After excitation, all three modes exhibit in-phase oscillation characteristic of CDW dynamics [Note S3], but $M_1$ is selectively amplified compared to $M_2$ and $M_3$ ($M_2 \approx M_3$), breaking the threefold rotational symmetry. The lattice simultaneously departs from equilibrium, forming nonequilibrium metastable states. While the distortion amplitude varies with laser parameters, the symmetry-breaking behavior is robust [Fig. S5]. Two representative transient structures at $t = 100$ fs and $170$ fs are shown in Fig. 2(b-c), where alternating V-V bond lengths along the $a$-axis reveal a dimer-like pattern. Relative to the ISD phase, the V-V bond distribution in these transient symmetry-breaking (TSB) states broadens, with a clear shift in the dominant peak [Fig. S6]. The simulated STM image of the Sb-terminated surface exhibits a unidirectional stripe-like modulation distinct from that of the ISD state [Fig. 2(d)].

The anisotropic phonon response is closely linked to momentum-resolved electron-phonon coupling[48,50,69]. As shown in Fig. S7, the charge population reveals suppressed carrier excitation along the $\Gamma - M_1$ direction. Consequently, the bond strength associated with $M_1$ mode differs from that of $M_2$ and $M_3$, breaking the threefold symmetry[70,71]. This degeneracy among the single-$\mathbf{Q}_M$ modes is lifted, rendering the excited-state potential energy surface (PES) asymmetric and preferentially driving lattice distortion along $M_1$ [Fig. 2(e)]. Interestingly, similar behavior is observed in monolayer *1T*-TiSe$_2$, where only one single-$\boldsymbol{Q}_M$ mode is excited under comparable laser conditions [72]. This parallel underscore the generality of symmetry-selective excitation as a mechanism for modulating CDW states. Notably, due to the geometric frustration intrinsic to Kagome lattice, such photoexcitation not only modifies ground-state symmetry but also induces nonlinear response with potentially topological character.

The tunability of the amplitudes of the three single-$\mathbf{Q}_M$ modes enables optical control of the relative intensities of CDW peaks observed in the STM measurement. Fourier transforms (FT) of simulated STM images for key structures, i.e., pristine, ISD and TSB phases are shown in Fig. S8, with their differences displayed in Fig. 2(f-g). In contrast to the pristine phase, the ISD structure exhibits clear 2 × 2 charge-order peaks ($Q_{2\times2}$) with nearly identical intensities, reflecting the ideal in-plane lattice symmetry in the absence of interlayer stacking [Fig. 2(f)]. Upon applying laser pulses polarized along different $\boldsymbol{Q}_M^\alpha$, the relative intensities of these $Q_{2\times2}$ peaks are modulated, in agreement with experimental observations[41], confirming the reliability of our simulated nonequilibrium dynamics.

Figure 3(a) displays the electronic structure of the TSB phase, showing a redistribution of the density of states (DOS) near the Fermi level ($E_F$) and evident band splitting relative to the ISD phase [Fig. S9]. A prominent bandgap opens at the *K(H)* points, indicating broken $\mathcal{C}_6$ rotational symmetry. This anisotropy is further manifested in band dispersion around the gap region, which is experimentally accessible via time- and angle-resolved photoemission spectroscopy (tr-ARPES) measurements[13,73,74]. To elucidate the origin of these electronic features, we employ a single-orbital tight-binding model for the V sublattice [Note S4]. The model reproduces the band evolution from the pristine to the ISD phase, including the formation of CDW gaps at and away from $E_F$ near the *M* point[69]. Importantly, introducing a lattice distortion corresponding to a single-$\boldsymbol{Q}_M$ mode lifts the degeneracy of the Dirac-like bands at *K(H)*, leading to a gap opening [Fig. S10], consistent with our DFT results. This confirms the microscopic mechanism of symmetry breaking and band reconstruction.

*Emergent ferrimagnetism and time-reversal symmetry breaking:*

Interesting, $\mathcal{T}$-breaking has been reported in ultrafast pump-probe experiments[41]. While full *ab initio* treatment of the spin (orbital)-phonon dynamics is computationally prohibitive, we investigate the magnetic nature of the TSB states via spin-polarized

DFT. Unlike the pristine and ISD phases, the TSB states exhibit spontaneous magnetization localized on V atoms. The magnetic anisotropy energy, defined as the energy difference upon rotating the magnetization direction[75], favors spin alignment along the laser polarization, i.e. *a*-axis, with energies 0.13 and 0.44 meV/cell lower than along the *c*- and *b*-axes, respectively. The induced spin moments display an alternating pattern along the *b*-axis, arising from unequal V-V bond length along *a*-axis, indicating strong coupling between lattice geometry and magnetic order [Fig. 3(b)]. This gives rise to a ferrimagnetic state with a net magnetic moment of ~0.5 $\mu_B$/cell at $t = 100$ fs, suggesting the dynamical stability of this nonequilibrium configuration. Similar phonon-driven magnetic excitations have been observed in other systems, where lattice distortions induce magnetic states distinct from equilibrium[76-78].

Symmetry analysis provides further insights. In pristine or ISD phase, the three sublattice sites form a frustrated triangular unit cell with isotropic exchange interactions (i.e., $t_1 = t_2 = t_3$) [Fig. 4(a-b)]. Exciting a single-$Q_M$ mode or introducing unequal $Q_M$-mode amplitudes result in anisotropic hopping, driving Fermi surface instabilities and the emergence of hidden states[4,73,79]. In an anisotropic Kagome lattice where one hopping path is inequivalent ($t_1' = t_2' \neq t_3'$) [Fig. 4(c)], frustration is relieved and the system transitions from paramagnetic to ferrimagnetic ground states[5,10,80,81]. Single-$Q_M$ excitation in the pristine phase also leads to a ferrimagnetic pattern, confirming geometry-driven magnetic excitations [Fig. S11(a-b)]. Notably, such ferrimagnetism is not induced by comparable uniaxial strain, indicating that the dimerization-like pattern of V–V bond distortions is essential [Fig. S11(c-d)]. Moreover, the persistence of net magnetization under interlayer stacking suggests applicability to more complex configurations [Fig. S12].

The observed ferrimagnetic state with $\mathcal{T}$-breaking implies the emergence of an AHE, as confirmed in our calculation. Figure 3(c) shows the Fermi surface and Berry curvature distribution in momentum space. Comparing to the ISD phase [Fig. S9(b)], the TSB state features pronounced Berry curvature "hot spots" near avoided crossings, which lead to an intrinsic AHE[82] [Fig. 3(d)]. According to the symmetry of magnetic point group m′m′m, only the $\sigma_{yz}$ component of the Hall tensor is allowed, where $\sigma_{zx}$ and $\sigma_{xy}$ are forbidden by symmetry operations including $\mathcal{T}\mathcal{C}_{2y}$, $\mathcal{T}\mathcal{C}_{2z}$, $\mathcal{T}\mathcal{M}_y$, $\mathcal{T}\mathcal{M}_z$, $\mathcal{C}_{2x}$, and $\mathcal{M}_x$. Despite a modest net moment, the calculated $\sigma_{yz}$ reaches ~174 S/cm near $E_F$, which can be further enhanced by electron doping, e.g., via tuning Rb content. This value exceeds the typical AHE conductivity in $Fe_3O_4$ (< 10 S/cm)[83], and approaches that of ferromagnets such as hcp Co [84](480 S/cm) and bcc Fe (750 S/cm)[85].

**Discussions**

In this work, we focus on the lattice-spin coupling that give rise to photo-induced magnetic excitation. Giving the ongoing debate over the existence of intrinsic magnetism and the complex many-body interactions involved[37,86,87], we assume the system begins in a paramagnetic state without delving further into this controversy. Unlike the flux-phase-like model derived from thermal equilibrium analysis[41], we show that: (1) polarization-dependent changes in CDW peak intensity could arise from either mode-selective excitation or electrostriction-induced anisotropic strain, or potentially from a combination of both; and (2) dimerization-like distortion of V–V bonds induce ferrimagnetism, establishing a spin-lattice coupling origin for $\mathcal{T}$-breaking. In the context of photoexcitation, the delicate balance of entangled orders is broken, a comprehensive discussion of symmetry breaking must consider both possible intrinsic properties and photo-induced effects.

Our results establish a connection between intrinsic AHE, photoinduced ferrimagnetic ordering, and the Berry curvature of electronic bands, which exhibit fundamentally different microscopic behavior compared to previous studies[17,18]. For instance, in $KV_3Sb_5$, skew scattering has been proposed as the origin of the enhanced AHE under an external magnetic field, attributed to the coexistence of Dirac quasiparticles and the frustrated sublattice, with no need for magnetic ordering [17,88,89]. Experimentally, both the skew scattering and intrinsic components of the AHE can be quantitatively determined by fitting to the equation $\sigma = \alpha \sigma_{xx0}^{-1} \sigma_{xx}^2 + b$, where $\alpha$ is the skew constant, $\sigma_{xx0}^{-1}$ is the residual resistivity, $\sigma_{xx}$ is the longitudinal conductivity and *b* represents the intrinsic anomalous Hall conductivity [90]. Upon photoinduced relief of frustration, the intrinsic component is expected to increase, while the skew scattering term might be suppressed. Up to now, an orbital magnetic moment of $\sim 0.02 \pm 0.01$ $\mu_B$ per vanadium triangle is supposed[40], but local moment spin magnetism does not show up in any thermal-equilibrium magnetic measurements[11,91]. Upon photoexcitation, net spin moment emerge and might be experimentally detected by the time-resolved pump-probe techniques such as magneto-optical Kerr effect (MOKE) measurement[92].

The demonstrated symmetry tunability underscores the feasibility of designing chiral architecture [Note S7] and electronic structures in Kagome materials. Due to the inherent geometrical frustration of the Kagome lattice, $A$V$_3$Sb$_5$ hosts a range of unique band features, including flat bands, Dirac cones and saddle-point van Hove singularities[73,74,93-96], that underpin diverse emergent phenomena such as topological order, superconductivity, and CDW formation[97,98]. By optically modulating lattice symmetry, the location and degeneracy of these electronic features can be selectively controlled, enabling dynamic tuning of the Fermi surface topology and band dispersion. This offers a powerful platform not only for validating theoretical predictions but also for developing optically reconfigurable quantum devices based on Kagome-lattice-derived electronic correlations.

## *Summary*

In conclusion, we identify a microscopic, optically driven pathway for symmetry breaking in RbV$_3$Sb$_5$, distinct from previously reported ground-state mechanisms. Selective enhancement of a single-$\mathbf{Q}_M$ phonon mode lifts rotational and time-reversal symmetries by inducing anisotropic electronic hopping and relieving geometric frustration. The resulting nonequilibrium ferrimagnetic phase exhibits a sizable intrinsic anomalous Hall effect arising from a net spin moment mediated by spin-orbit coupling. In contrast to prior interpretations invoking orbital currents or extrinsic perturbations, our findings provide direct evidence for a spin-driven mechanisms.

These results highlight the fundamental role of light in accessing hidden magnetic and topological orders and open new avenues for ultrafast control of symmetry, magnetism, and electronic topology in frustrated quantum materials.

*Acknowledgement*

M.-X. G. acknowledges partial financial support from the National Key Research and Development Program of China (No. 2024YFA1207800), National Natural Science Foundation of China (No. 12304536), and the start-up funding of Beijing Institute of Technology. Y.-G. Y. acknowledges partial financial support from the National Key Research and Development Program of China (No. 2020YFA0308800), National Natural Science Foundation of China (No. 12321004 and No. 12234003). X.-D. Z. acknowledges support from the National Natural Science Foundation of China (Grant No. 12304066) and the Basic Research Program of Jiangsu (Grant No. BK20230684).

[1] Y. Wang, H. Wu, G. T. McCandless, J. Y. Chan, and M. N. Ali, *Quantum states and intertwining phases in kagome materials,* Nature Reviews Physics **5**, 635 (2023).
[2] A. P. Ramirez, *Strongly geometrically frustrated magnets,* Annu. Rev. Mater. Sci. **24**, 453 (1994).
[3] R. Moessner and A. Ramirez, *Geometrical frustration,* Phys. Today **59**, 24 (2006).
[4] M. L. Kiesel and R. Thomale, *Sublattice interference in the kagome Hubbard model,* Phys. Rev. B **86**, 121105(R) (2012).
[5] Y. Imai, N. Kawakami, and H. Tsunetsugu, *Low-energy excitations of the Hubbard model on the Kagomé lattice,* Phys. Rev. B **68**, 195103 (2003).
[6] A. Simonov and A. L. Goodwin, *Designing disorder into crystalline materials,* Nat Rev Chem **4**, 657 (2020).
[7] J. Wang, M. Spitaler, Y. S. Su, K. M. Zoch, C. Krellner, P. Puphal, S. E. Brown, and A. Pustogow, *Controlled Frustration Release on the Kagome Lattice by Uniaxial-Strain Tuning,* Phys. Rev. Lett. **131**, 256501 (2023).
[8] X. Fabreges, S. Petit, I. Mirebeau, S. Pailhes, L. Pinsard, A. Forget, M. T. Fernandez-Diaz, and F. Porcher, *Spin-lattice coupling, frustration, and magnetic order in multiferroic RMnO3,* Phys. Rev. Lett. **103**, 067204 (2009).
[9] J. Ma *et al.*, *Static and Dynamical Properties of the Spin-1/2 Equilateral Triangular-Lattice Antiferromagnet $Ba_3CoSb_2O_9$,* Phys. Rev. Lett. **116**, 087201 (2016).
[10] C. Xu, S. Wu, G.-X. Zhi, G. Cao, J. Dai, C. Cao, X. Wang, and H.-Q. Lin, *Altermagnetic ground state in distorted Kagome metal CsCr3Sb5,* Nat. Commun. **16**, 3114 (2025).
[11] B. R. Ortiz *et al.*, *New kagome prototype materials: discovery of KV3Sb5, RbV3Sb5, and CsV3Sb5,* Physical Review Materials **3**, 094407 (2019).
[12] Z. X. Wang *et al.*, *Unconventional charge density wave and photoinduced lattice symmetry change*

*in the kagome metal CsV3Sb5 probed by time-resolved spectroscopy,* Phys. Rev. B **104**, 165110 (2021).

[13] B. R. Ortiz *et al.*, *CsV3Sb5: A Z2 Topological Kagome Metal with a Superconducting Ground State,* Phys. Rev. Lett. **125**, 247002 (2020).

[14] C. Mielke, 3rd *et al.*, *Time-reversal symmetry-breaking charge order in a kagome superconductor,* Nature **602**, 245 (2022).

[15] C. W. L. Yu, Y. Zhang, M. Sander, S. Ni, Z. Lu, S. Ma,, Z. Z. Z. Wang, H. Chen, K. Jiang, Y. Zhang, H. Yang,, X. D. F. Zhou, S. L. Johnson, M. J. Graf, J. Hu, H. J. Gao, and a. Z. Zhao, *Evidence of a hidden flux phase in the topological kagome metal CsV3Sb5,* preprint, arXiv:2107.10714, (2021).

[16] R. Khasanov *et al.*, *Time-reversal symmetry broken by charge order in CsV3Sb5,* Physical Review Research **4** (2022).

[17] S.-Y. Yang *et al.*, *Giant, unconventional anomalous Hall effect in the metallic frustrated magnet candidate, KV3Sb5,* Science Advances **6**, eabb6003 (2020).

[18] F. H. Yu, T. Wu, Z. Y. Wang, B. Lei, W. Z. Zhuo, J. J. Ying, and X. H. Chen, *Concurrence of anomalous Hall effect and charge density wave in a superconducting topological kagome metal,* Phys. Rev. B **104**, L041103 (2021).

[19] Y. Xu, Z. Ni, Y. Liu, B. R. Ortiz, Q. Deng, S. D. Wilson, B. Yan, L. Balents, and L. Wu, *Three-state nematicity and magneto-optical Kerr effect in the charge density waves in kagome superconductors,* Nature Phys. **18**, 1470 (2022).

[20] X. Wei *et al.*, *Three-dimensional hidden phase probed by in-plane magnetotransport in kagome metal CsV3Sb5 thin flakes,* Nat. Commun. **15**, 5038 (2024).

[21] H. Deng *et al.*, *Chiral kagome superconductivity modulations with residual Fermi arcs,* Nature **632**, 775 (2024).

[22] C. Guo *et al.*, *Switchable chiral transport in charge-ordered kagome metal CsV3Sb5,* Nature **611**, 461 (2022).

[23] Y. X. Jiang *et al.*, *Unconventional chiral charge order in kagome superconductor KV3Sb5,* Nat. Mater. **20**, 1353 (2021).

[24] Z. J. Cheng *et al.*, *Broken symmetries associated with a Kagome chiral charge order,* Nat. Commun. **16**, 3782 (2025).

[25] H. D. Scammell, J. Ingham, T. Li, and O. P. Sushkov, *Chiral excitonic order from twofold van Hove singularities in kagome metals,* Nat Commun **14**, 605 (2023).

[26] X. Feng, K. Jiang, Z. Wang, and J. Hu, *Chiral flux phase in the Kagome superconductor AV3Sb5,* Sci. Bull. **66**, 1384 (2021).

[27] H.-Y. Ma, J.-X. Yin, M. Zahid Hasan, and J. Liu, *Magnetic and charge instabilities in vanadium-based topological kagome metals,* Phys. Rev. B **106**, 155125 (2022).

[28] M. H. Christensen, T. Birol, B. M. Andersen, and R. M. Fernandes, *Loop currents in AV3Sb5 kagome metals: Multipolar and toroidal magnetic orders,* Phys. Rev. B **106**, 144504 (2022).

[29] P. Negi, K. Medhi, A. Pancholi, and S. Roychowdhury, *Magnetic Kagome materials: bridging fundamental properties and topological quantum applications,* Mater Horiz **12**, 4510 (2025).

[30] J. X. Yin, B. Lian, and M. Z. Hasan, *Topological kagome magnets and superconductors,* Nature **612**, 647 (2022).

[31] P. Wu *et al.*, *Unidirectional electron–phonon coupling in the nematic state of a kagome superconductor,* Nature Phys. **19**, 1143 (2023).

[32] L. Nie *et al.*, *Charge-density-wave-driven electronic nematicity in a kagome superconductor,* Nature **604**, 59 (2022).

[33] H. Chen *et al.*, *Roton pair density wave in a strong-coupling kagome superconductor,* Nature **599**, 222 (2021).

[34] S. D. Wilson and B. R. Ortiz, *AV3Sb5 kagome superconductors,* Nature Reviews Materials **9**, 420 (2024).

[35] H. Li, H. Zhao, B. R. Ortiz, T. Park, M. Ye, L. Balents, Z. Wang, S. D. Wilson, and I. Zeljkovic, *Rotation symmetry breaking in the normal state of a kagome superconductor KV3Sb5,* Nature Phys. **18**, 265 (2022).

[36] T. Park, M. Ye, and L. Balents, *Electronic instabilities of kagome metals: Saddle points and Landau theory,* Phys. Rev. B **104**, 035142 (2021).

[37] C. Guo *et al.*, *Correlated order at the tipping point in the kagome metal CsV3Sb5,* Nat Phys **20**, 579 (2024).

[38] D. R. Saykin *et al.*, *High Resolution Polar Kerr Effect Studies of CsV3Sb5: Tests for Time-Reversal Symmetry Breaking below the Charge-Order Transition,* Phys. Rev. Lett. **131**, 016901 (2023).

[39] H. Li *et al.*, *No observation of chiral flux current in the topological kagome metal CsV3Sb5,* Phys. Rev. B **105**, 045102 (2022).

[40] W. Liège *et al.*, *Search for orbital magnetism in the kagome superconductor CsV3Sb5 using neutron diffraction,* Phys. Rev. B **110**, 195109 (2024).

[41] Y. Xing *et al.*, *Optical manipulation of the charge-density-wave state in RbV3Sb5,* Nature **631**, 60 (2024).

[42] M. H. Christensen, T. Birol, B. M. Andersen, and R. M. Fernandes, *Theory of the charge density wave in AV3Sb5 kagome metals,* Phys. Rev. B **104**, 214513 (2021).

[43] H. Tan, Y. Liu, Z. Wang, and B. Yan, *Charge Density Waves and Electronic Properties of Superconducting Kagome Metals,* Phys. Rev. Lett. **127**, 046401 (2021).

[44] C. Wang, S. Liu, H. Jeon, and J.-H. Cho, *Origin of charge density wave in the layered kagome metal CsV3Sb5,* Phys. Rev. B **105**, 045135 (2022).

[45] B. Zhang, H. Tan, B. Yan, C. Xu, and H. Xiang, *Atomistic Origin of Diverse Charge Density Wave States in $CsV_3Sb_5$,* Phys. Rev. Lett. **132**, 096101 (2024).

[46] G. He *et al.*, *Anharmonic strong-coupling effects at the origin of the charge density wave in CsV3Sb5,* Nat. Commun. **15**, 1895 (2024).

[47] G. Liu *et al.*, *Observation of anomalous amplitude modes in the kagome metal CsV3Sb5,* Nat. Commun. **13**, 3461 (2022).

[48] D. Azoury *et al.*, *Direct observation of the collective modes of the charge density wave in the kagome metal CsV3Sb5,* Proc Natl Acad Sci U S A **120**, e2308588120 (2023).

[49] J. Yu *et al.*, *All-optical manipulation of charge density waves in kagome metal CsV3Sb5,* Phys. Rev. B **107**, 174303 (2023).

[50] Y. Xie *et al.*, *Electron-phonon coupling in the charge density wave state of CsV3Sb5,* Phys. Rev. B **105**, L140501 (2022).

[51] See Supplemental Material at xx for details about the theoretical methods and other discussion.

[52] G. Kresse and D. Joubert, *From ultrasoft pseudopotentials to the projector augmented-wave method,* Phys. Rev. B **59**, 1758 (1999).

[53] G. Kresse and J. Furthmüller, *Efficiency of ab-initio total energy calculations for metals and semiconductors using a plane-wave basis set,* Comp. Mater. Sci. **6**, 15 (1996).

[54] G. Kresse and J. Furthmüller, *Efficient iterative schemes for ab initio total-energy calculations using a plane-wave basis set,* Phys. Rev. B **54**, 11169 (1996).

[55] J. P. Perdew, K. Burke, and M. Ernzerhof, *Generalized gradient approximation made simple,* Phys. Rev. Lett. **77**, 3865 (1996).

[56] A. A. Mostofi, J. R. Yates, Y.-S. Lee, I. Souza, D. Vanderbilt, and N. Marzari, *wannier90: A tool for obtaining maximally-localised Wannier functions,* Computer physics communications **178**, 685 (2008).

[57] S. Grimme, J. Antony, S. Ehrlich, and H. Krieg, *A consistent and accurate ab initio parametrization of density functional dispersion correction (DFT-D) for the 94 elements H-Pu,* J. Chem. Phys. **132**, 154104 (2010).

[58] A. Togo and I. Tanaka, *First principles phonon calculations in materials science,* Scripta Materialia **108**, 1 (2015).

[59] C. Lian, M. Guan, S. Hu, J. Zhang, and S. Meng, *Photoexcitation in Solids: First-Principles Quantum Simulations by Real-Time TDDFT,* Adv. Theory. Simul. **1**, 1800055 (2018).

[60] K. Choudhary, K. F. Garrity, C. Camp, S. V. Kalinin, R. Vasudevan, M. Ziatdinov, and F. Tavazza, *Computational scanning tunneling microscope image database,* Sci Data **8**, 57 (2021).

[61] V. Wang, N. Xu, J.-C. Liu, G. Tang, and W.-T. Geng, *VASPKIT: A user-friendly interface facilitating high-throughput computing and analysis using VASP code,* Computer Physics Communications **267**, 108033 (2021).

[62] M. Guan, D. Chen, S. Hu, H. Zhao, P. You, and S. Meng, *Theoretical Insights into Ultrafast Dynamics in Quantum Materials,* Ultrafast Science **2022**, 9767251 (2022).

[63] Z.-M. Y. Zeying Zhang, Gui-Bin Liu, and Yugui Yao, *A phonon irreducible representations calculator,* arXiv preprint arXiv:2201.11350 (2022).

[64] Z. Zhang, Z.-M. Yu, G.-B. Liu, and Y. Yao, *MagneticTB: A package for tight-binding model of magnetic and non-magnetic materials,* Computer Physics Communications **270** (2022).

[65] F. Jin, W. Ren, M. Tan, M. Xie, B. Lu, Z. Zhang, J. Ji, and Q. Zhang, *pi Phase Interlayer Shift and Stacking Fault in the Kagome Superconductor CsV3Sb5,* Phys. Rev. Lett. **132**, 066501 (2024).

[66] Q. Wu, S. Zhang, H.-F. Song, M. Troyer, and A. A. Soluyanov, *WannierTools: An open-source software package for novel topological materials,* Computer Physics Communications **224**, 405 (2018).

[67] J. Ishioka, Y. H. Liu, K. Shimatake, T. Kurosawa, K. Ichimura, Y. Toda, M. Oda, and S. Tanda, *Chiral charge-density waves,* Phys. Rev. Lett. **105**, 176401 (2010).

[68] S. Y. Xu *et al.*, *Spontaneous gyrotropic electronic order in a transition-metal dichalcogenide,* Nature **578**, 545 (2020).

[69] H. Luo *et al.*, *Electronic nature of charge density wave and electron-phonon coupling in kagome superconductor KV3Sb5,* Nat. Commun. **13**, 273 (2022).

[70] X. B. Li, X. Q. Liu, X. Liu, D. Han, Z. Zhang, X. D. Han, H. B. Sun, and S. B. Zhang, *Role of electronic excitation in the amorphization of Ge-Sb-Te alloys,* Phys. Rev. Lett. **107**, 015501 (2011).

[71] Z. N. Scheller, S. Mehrparvar, and G. Haberhauer, *Light-Induced Increase in Bond Strength horizontal line from Chalcogen Bond to Three-Electron sigma Bond upon Excitation,* J. Am. Chem. Soc. **147**, 6249 (2025).

[72] Z. Nie, Y. Wang, D. Chen, and S. Meng, *Unraveling Hidden Charge Density Wave Phases in 1T−TiSe2,* Phys. Rev. Lett. **131**, 196401 (2023).

[73] Y. Hu *et al.*, *Rich nature of Van Hove singularities in Kagome superconductor CsV3Sb5,* Nat. Commun. **13**, 2220 (2022).

[74] M. Kang *et al.*, *Twofold van Hove singularity and origin of charge order in topological kagome superconductor CsV3Sb5,* Nature Phys. **18**, 301 (2022).

[75] P. Larson, I. I. Mazin, and D. A. Papaconstantopoulos, *Calculation of magnetic anisotropy energy*

*inSmCo5,* Phys. Rev. B **67**, 214405 (2003).

[76] T. Kahana, D. A. Bustamante Lopez, and D. M. Juraschek, *Light-induced magnetization from magnonic rectification,* Science Advances **10**, eado0722 (2024).

[77] S. Zhang, Y. Pei, S. Hu, N. Wu, D.-Q. Chen, C. Lian, and S. Meng, *Light-induced phonon-mediated magnetization in monolayer MoS2,* Chinese Physics Letters **40**, 077502 (2023).

[78] M. Fechner, A. Sukhov, L. Chotorlishvili, C. Kenel, J. Berakdar, and N. A. Spaldin, *Magnetophononics: Ultrafast spin control through the lattice,* Physical Review Materials **2**, 064401 (2018).

[79] Y. L. Lin and F. Nori, *Quantum interference on the kagome-acute lattice,* Phys. Rev. B **50**, 15953 (1994).

[80] A. Yamada, K. Seki, R. Eder, and Y. Ohta, *Mott transition and ferrimagnetism in the Hubbard model on the anisotropic kagome lattice,* Phys. Rev. B **83**, 195127 (2011).

[81] L. O. Lima, A. R. Medeiros-Silva, R. R. dos Santos, T. Paiva, and N. C. Costa, *Magnetism and metal-insulator transitions in the anisotropic kagome lattice,* Phys. Rev. B **108**, 235163 (2023).

[82] N. Nagaosa, J. Sinova, S. Onoda, A. H. MacDonald, and N. P. Ong, *Anomalous Hall effect,* Rev. Mod. Phys. **82**, 1539 (2010).

[83] A. Fernández-Pacheco, J. M. De Teresa, J. Orna, L. Morellon, P. A. Algarabel, J. A. Pardo, and M. R. Ibarra, *Universal scaling of the anomalous Hall effect inFe3O4epitaxial thin films,* Phys. Rev. B **77** (2008).

[84] X. Wang, D. Vanderbilt, J. R. Yates, and I. Souza, *Fermi-surface calculation of the anomalous Hall conductivity,* Phys. Rev. B **76** (2007).

[85] Y. Yao, L. Kleinman, A. H. MacDonald, J. Sinova, T. Jungwirth, D. S. Wang, E. Wang, and Q. Niu, *First principles calculation of anomalous Hall conductivity in ferromagnetic bcc Fe,* Phys. Rev. Lett. **92**, 037204 (2004).

[86] D. Karmakar *et al.*, *Magnetism in $AV_3Sb_5$ ($A=Cs,Rb,K$): Complex landscape of dynamical magnetic textures,* Phys. Rev. B **108** (2023).

[87] M. N. Hasan *et al.*, *Magnetism in $AV_3Sb_5$ (A=Cs, Rb, and K): Origin and Consequences for the Strongly Correlated Phases,* Phys. Rev. Lett. **131**, 196702 (2023).

[88] H. Ishizuka and N. Nagaosa, *Large anomalous Hall effect and spin Hall effect by spin-cluster scattering in the strong-coupling limit,* Phys. Rev. B **103**, 235148 (2021).

[89] H. Ishizuka and N. Nagaosa, *Spin chirality induced skew scattering and anomalous Hall effect in chiral magnets,* **4**, eaap9962 (2018).

[90] Y. Tian, L. Ye, and X. Jin, *Proper scaling of the anomalous Hall effect,* Phys. Rev. Lett. **103**, 087206 (2009).

[91] E. M. Kenney, B. R. Ortiz, C. Wang, S. D. Wilson, and M. J. Graf, *Absence of local moments in the kagome metal KV(3)Sb(5)as determined by muon spin spectroscopy,* J Phys Condens Matter **33** (2021).

[92] G. P. Zhang, W. Hübner, G. Lefkidis, Y. Bai, and T. F. George, *Paradigm of the time-resolved magneto-optical Kerr effect for femtosecond magnetism,* Nature Phys. **5**, 499 (2009).

[93] K. Zeng, Z. Wang, K. Jiang, and Z. Wang, *Electronic structure of AV3Sb5 kagome metals,* Phys. Rev. B **111**, 235114 (2025).

[94] Y.-P. Lin and R. M. Nandkishore, *Complex charge density waves at Van Hove singularity on*

*hexagonal lattices: Haldane-model phase diagram and potential realization in the kagome metals AV3Sb5 (A=K, Rb, Cs),* Phys. Rev. B **104**, 045122 (2021).

[95] J. Zhao, W. Wu, Y. Wang, and S. A. Yang, *Electronic correlations in the normal state of the kagome superconductor KV3Sb5,* Phys. Rev. B **103**, L241117 (2021).

[96] T. Kato *et al.*, *Three-dimensional energy gap and origin of charge-density wave in kagome superconductor KV3Sb5,* Communications Materials **3** (2022).

[97] M. M. Denner, R. Thomale, and T. Neupert, *Analysis of Charge Order in the Kagome Metal AV3Sb5 (A=K,Rb,Cs),* Phys. Rev. Lett. **127**, 217601 (2021).

[98] H. Li *et al.*, *Observation of Unconventional Charge Density Wave without Acoustic Phonon Anomaly in Kagome Superconductors AV3Sb5 (A=Rb, Cs),* Phy. Rev. X **11**, 031050 (2021).

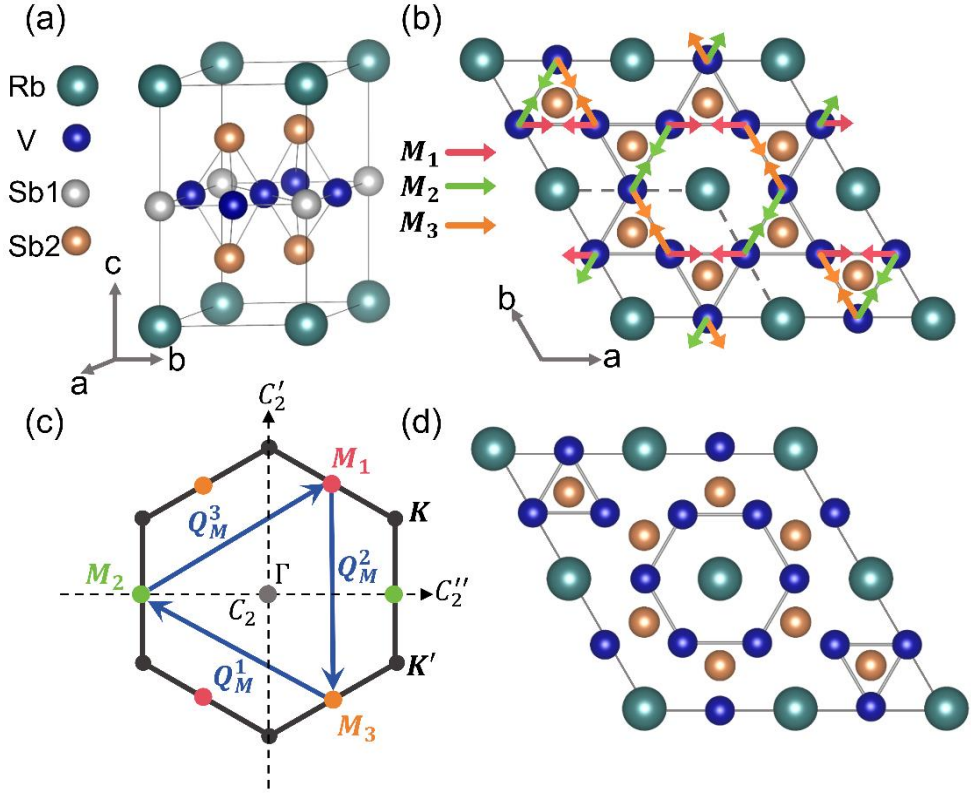

FIG. 1. (a-b) Crystal structures of RbV$_3$Sb$_5$ in the pristine phase. Sb atoms within the V plane are labeled as Sb1, while others are labeled as Sb2. The imaginary phonon modes at the three $M$ points are shown in a 2×2×1 superlattice, reflecting the $\pi$ phase in the phonon dynamical matrix eigenvector. (c) Hexagonal Brillouin zone of RbV$_3$Sb$_5$, showing the $M_\alpha$ points connected by three nesting vectors $\boldsymbol{Q}_M^\alpha$. (d) Crystal structure of the ISD phase, characterized by lattice distortion described with $M_1=M_2=M_3$ and $M_1 M_2 M_3 > 0$, corresponding to the breathing motion.

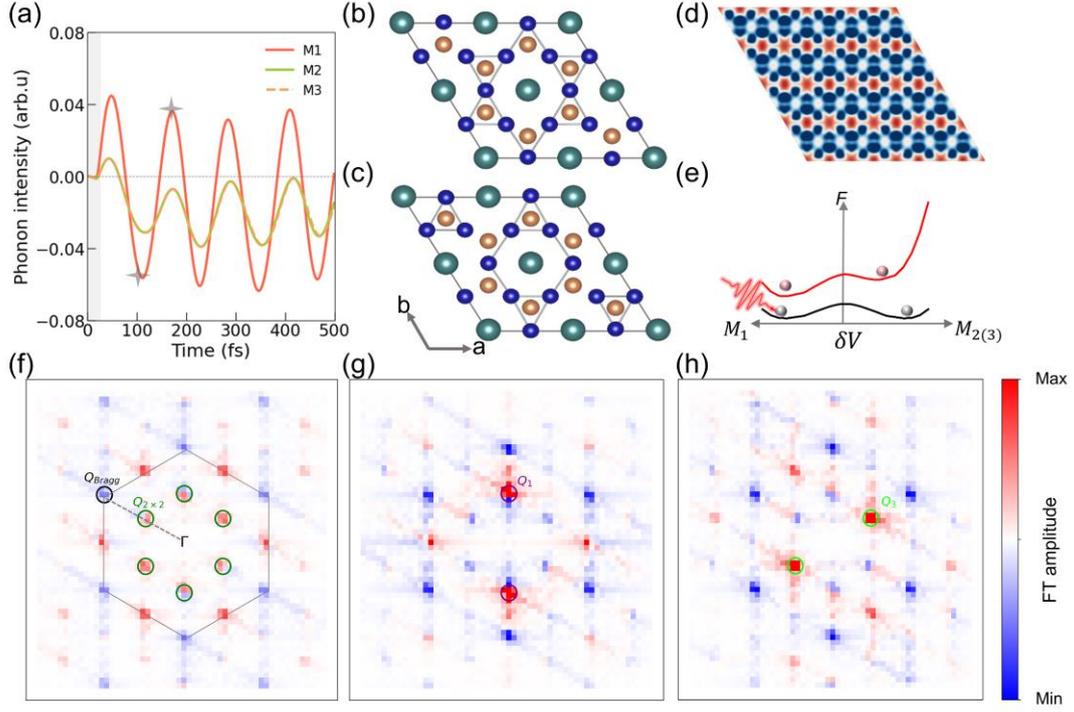

FIG. 2. (a) Projected vibrational amplitude of the three single-$Q_M$ phonon modes driven by an ultrafast laser pulse (gray dashed region), polarized along the *a*-axis with a fluence of $0.3\ \mathrm{mJ/cm^2}$. (b-c) Snapshots of transient atomic structures along the dynamical trajectory at (b) $t = 100$ fs and (c) $t = 170$ fs, as marked in (a) by stars. (d) Simulated STM image corresponding to the structure in (b). (e) Schematic of the potential-energy landscape along the three vibrational modes. The upper (lower) panel represents laser-excited (equilibrium) PESs. (f-h) Difference maps of the Fourier transform (FT) of STM images: (f) ISD state minus pristine; (g) transient state at $t = 100$ fs minus ISD, under laser polarized along $\boldsymbol{Q}_M^1$; (h) same as (g) but with polarization along $\boldsymbol{Q}_M^3$.

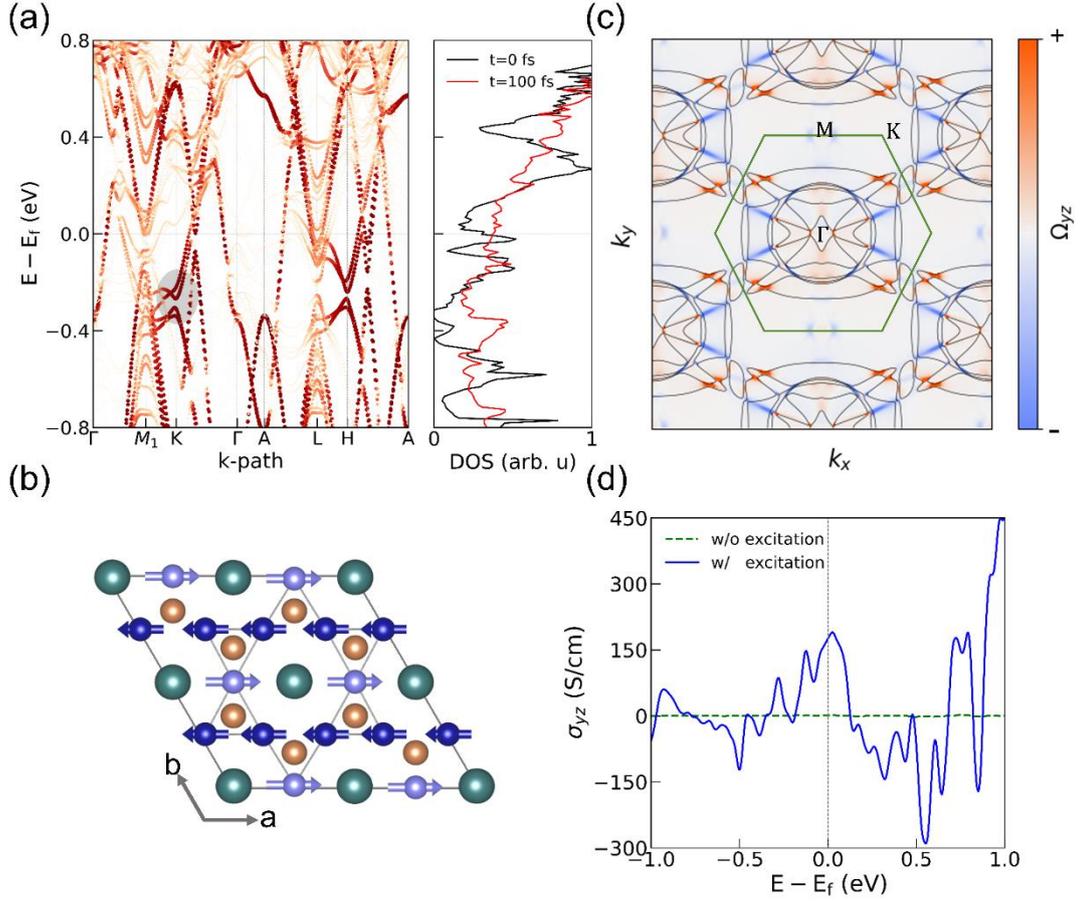

FIG. 3. Electronic and magnetic properties of the TSB state at $t = 100$ fs. (a) Left: Band structure of the TSB state. Right: Comparison of the total DOS between the TSB state and ISD state. (b) Spatial distribution of spin momentum on V atoms in real-space. (c) Fermi surface (black lines) and Berry curvature distribution in momentum space. The green hexagon outlines the first Brillouin zone. (d) Calculated anomalous Hall conductivity for the TSB (laser-excited) and ISD (equilibrium) phases.

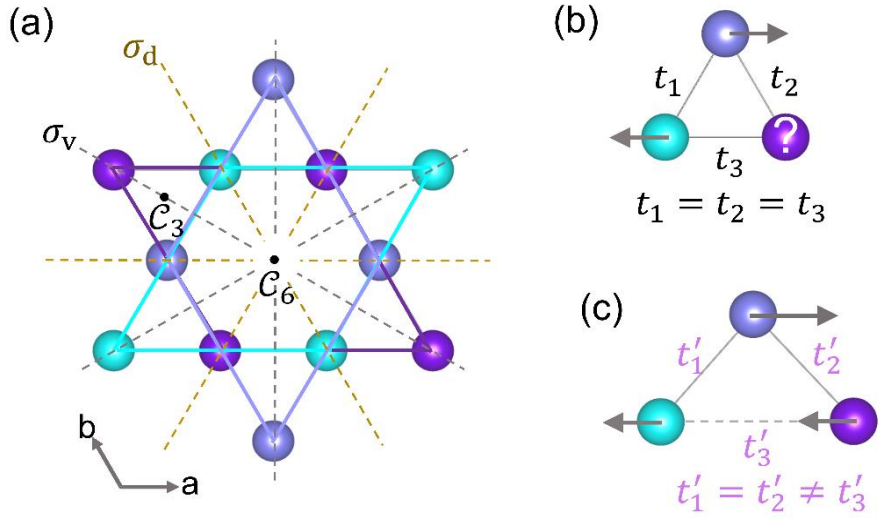

FIG. 4. (a) Top view of the monolayer V Kagome lattice, with key symmetry elements indicated. The three Kagome sublattices are color coded. (b) Geometrical frustration in the undistorted (isotropic) Kagome lattice with uniform nearest-neighbor hopping integrals $t_i$. (c) Distorted Kagome lattice with anisotropic hopping (e.g., $t'_1 = t'_2 \neq t'_3$), leading to frustration lifting and spin polarization.

Supplementary Information for

# Ultrafast symmetry modulation and induced magnetic excitation in the Kagome metal RbV$_3$Sb$_5$

Mengxue Guan *et al.*

**Note S1: Calculation methods**

Density functional theory (DFT) calculations were performed using the projector augmented wave (PAW) [1] method as implemented in the Vienna *Ab initio* Simulation Package (VASP)[2,3]. The plane wave energy cutoff was set to 300 eV. Exchange-correlation interactions were treated using the Perdew-Burke-Ernzerhof (PBE)[4] functional together with the zero damping DFT-D3 van der Waals correction[5]. Atomic positions were fully relaxed until the residual forces on all atoms were below $1 \times 10^{-6}$ eV/Å. For calculations of the $2 \times 2 \times 1$ supercell of RbV$_3$Sb$_5$, the Brillouin zone was sampled using a $6 \times 6 \times 8$ $k$-point mesh. Spin-orbit coupling (SOC) was included in band structure and magnetic property calculations. Magnetic configurations were determined self-consistently by total energy minimization. Although initial magnetic moments were aligned ferromagnetically, the system relaxed into a ferrimagnetic configuration due to symmetry constraints. The magnetocrystalline anisotropy energy calculations, a denser $9 \times 9 \times 11$ $k$-mesh was used with symmetry operations disabled.

To evaluate the effect of lattice distortions on the electronic structure, effective band structures were unfolded to the primitive Brillouin zone using the VASPKIT code[6]. The tight-binding models for non-magnetic configurations were constructed using the MagneticTB package[7]. Phonon dispersions were calculated using the finite displacement method as implemented in Phonopy[8]. A $3 \times 3 \times 2$ supercell with a $6 \times 6 \times 6$ mesh was used for the pristine phase, and a $2 \times 2 \times 1$ supercell with a $3 \times 3 \times 4$ mesh for the ISD phase. Symmetry classification and irreducible representations of phonon modes at arbitrary $k$-points were obtained using the PhononIrep package[9].

Constant-current STM images for Sb-terminated surfaces were simulated using a three-unit-cell-thick slab model within the Tersoff-Hamman method[10]. The slabs were unrelaxed to retain bulk properties, and a vacuum spacing of ~10 Å was included. The local density of states (LDOS) was calculated at a height of ~2 Å above the surface, integrated over an energy window of $-200$ meV relative to the bias voltage.

The anomalous Hall conductivity of the laser-induced magnetic state was evaluated using the WannierTools package[11] on a dense $201 \times 201 \times 201$ $\mathbf{k}$-mesh. A total of 240 maximally localized Wannier functions, constructed from 120 $d$ orbitals of $V$ and 120 $p$ orbitals from Sb atoms, were generated using WANNIER90 [12].

The TDDFT-MD simulations were performed with the Time Dependent Ab-initio Package (TDAP) as implemented in SIESTA[13,14]. Numerical atomic orbitals with double zeta polarization (DZP) were employed as the basis set, and electron-nuclear interactions were described by Troullier-Martins pseudopotentials. A real-space grid equivalent to a 250 Ry plane-wave cutoff was adopted. The Brillouin zone was sampled using a 3×3×3 $\mathbf{k}$-mesh for the $2 \times 2 \times 1$ supercell of RbV$_3$Sb$_5$ and the PBE-D3 zero damping van der Waals correction is applied. The coupling between atomic and electronic motions is governed by the Ehrenfest approximation. During dynamic simulations, the evolving time step was set to 0.05 fs for both electrons and ions in a micro-canonical ensemble. To simulate the laser-matter interaction, a velocity gauge was adopted with laser fields described by $E(t) = E_0 \cos(\omega t) \exp\left[-\frac{(t-t_0)^2}{2\sigma^2}\right]$, where the pulses are centered at $t_0 = 20$ fs with photon energy $\hbar\omega = 1.21$ eV (wavelength $\lambda$ =1025 nm) and the width $\sigma$ = 5 fs. The peak field $E_0$ ranged from 0.02 to 0.04 V/Å, polarized along the in-plane V-V nearest-neighbor directions. The laser fluence $F = \int_0^T I(t)dt$, where $I(t) = \frac{1}{2}c\varepsilon_0 n|E(t)|^2$ is the transient laser intensity, $T$ is the laser duration, $c$ is the vacuum velocity of light. For RbV$_3$Sb$_5$, the measured refractive index is $n$= 3.16 at 5 K [Fig. S1(b)]. When $E_0 = 0.04$ V/Å, $F = 0.3$ mJ/cm$^2$, closely matching the experimental threshold ($F_c \sim 0.2$ mJ/cm$^2$) [15].

The momentum-resolved distribution of photoexcited carriers is calculated by projecting the time-dependent Kohn-Sham wavefunction $|\psi_{n,k}(t)\rangle$ onto the basis of the ground-state wavefunctions $|\psi_{n',k}(0)\rangle$,

$$\eta_k(t) = \sum_n \sum_{n'}^{CB} |\langle \psi_{n',k}(0)|\psi_{n,k}(t)\rangle|^2,$$

where $n$ sums over the band indices and $n'$ suns over all conduction bands. The light-induced photocurrent can be obtained by

$$J(t) = \frac{1}{2i} \int_\Omega dr \sum_i \{\psi_i^*(r,t)\nabla\psi_i(r,t) - \psi_i(r,t)\nabla\psi_i^*(r,t)\},$$

where $\psi_i$ denotes the time-evolved wave function of the $i$th occupied eigenstate. Circularly polarized lasers incident along the out-of-plane direction of the chiral CDW were modeled with $E(t) = E_0\{\cos(\omega t)\vec{x} + \cos(\omega t \pm \frac{\pi}{2})\vec{y}\}\exp\left[-\frac{(t-t_0)^2}{2\sigma^2}\right]$ [Note S7], where the pulses are centered at $t_0 = 20$ fs with $\hbar\omega = 1.21$ eV ($\lambda = 1025$ nm), $\sigma = 4$ fs and $E_0 = 0.015$ V/Å [Fig. S13(a-b)].

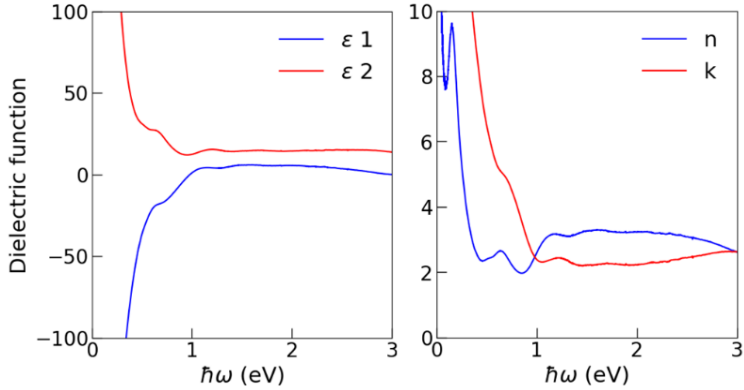

FIG. S1. Dielectric functions of RbV$_3$Sb$_5$ at 5 K.

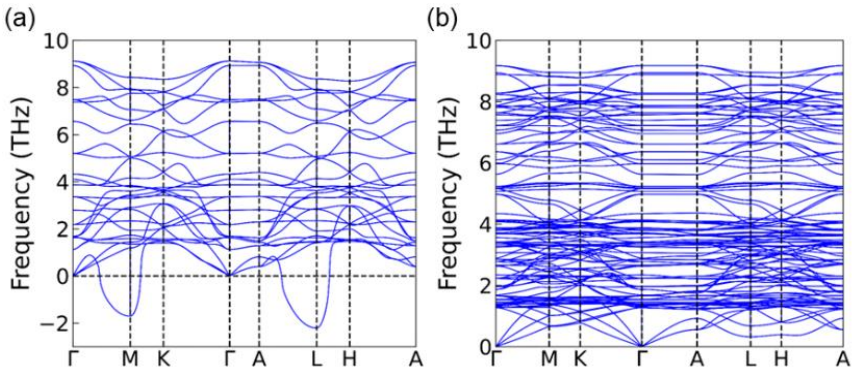

FIG. S2. Phonon dispersion of RbV$_3$Sb$_5$ in (a) pristine phase and (b) ISD phase.

**TABLE. S1.** Irreducible representations of the phonon modes at the $\Gamma$ point of the ISD phase. The degeneracy of phonon modes with identical frequencies is denoted by 'dim'.

| Band index | dim | Frequency(cm$^{-1}$) | Irep |
|---|---|---|---|
| 1, 2 | 2 | 0.00 | $E_{1u}, \Gamma_5^-(1)$ |
| 3 | 1 | 0.00 | $A_{2u}, \Gamma_2^-(1)$ |
| 4 | 1 | 42.30 | $B_{2g}, \Gamma_4^+(1)$ |
| 5, 6 | 2 | 42.52 | $E_{1g}, \Gamma_5^+(1)$ |
| 7 | 1 | 43.50 | $B_{1u}, \Gamma_3^-(1)$ |
| 8, 9 | 2 | 44.41 | $E_{1u}, \Gamma_5^-(1)$ |
| 10, 11 | 2 | 45.38 | $E_{2u}, \Gamma_6^-(1)$ |
| 12 | 1 | 46.53 | $B_{2u}, \Gamma_4^-(1)$ |
| 13 | 1 | 46.65 | $A_{2u}, \Gamma_2^-(1)$ |
| 14, 15 | 2 | 48.38 | $E_{1u}, \Gamma_5^-(1)$ |
| 16 | 1 | 51.08 | $A_{2g}, \Gamma_2^+(1)$ |
| 17, 18 | 2 | 51.48 | $E_{1u}, \Gamma_5^-(1)$ |
| 19, 20 | 2 | 55.36 | $E_{2g}, \Gamma_6^+(1)$ |
| 21 | 1 | 55.49 | $A_{2u}, \Gamma_2^-(1)$ |
| 22 | 1 | 56.17 | $B_{1u}, \Gamma_3^-(1)$ |
| 23, 24 | 2 | 62.32 | $E_{1u}, \Gamma_5^-(1)$ |
| 25, 26 | 2 | 65.33 | $E_{2u}, \Gamma_6^-(1)$ |
| 27 | 1 | 65.49 | $A_{2u}, \Gamma_2^-(1)$ |
| 28 | 1 | 74.30 | $A_{1u}, \Gamma_1^-(1)$ |
| 29, 30 | 2 | 74.45 | $E_{1g}, \Gamma_5^+(1)$ |
| 31, 32 | 2 | 78.05 | $E_{2u}, \Gamma_6^-(1)$ |
| 33, 34 | 2 | 87.66 | $E_{2g}, \Gamma_6^+(1)$ |
| 35 | 1 | 88.82 | $A_{2u}, \Gamma_2^-(1)$ |
| 36, 37 | 2 | 92.86 | $E_{1u}, \Gamma_5^-(1)$ |
| 38 | 1 | 95.13 | $A_{2u}, \Gamma_2^-(1)$ |
| 39, 40 | 2 | 95.23 | $E_{2u}, \Gamma_6^-(1)$ |
| 41 | 1 | 96.45 | $A_{1g}, \Gamma_1^+(1)$ |
| 42 | 1 | 102.26 | $B_{1u}, \Gamma_3^-(1)$ |
| 43 | 1 | 108.46 | $B_{2u}, \Gamma_4^-(1)$ |
| 44 | 1 | 109.08 | $B_{1g}, \Gamma_3^+(1)$ |
| 45, 46 | 2 | 110.99 | $E_{1g}, \Gamma_5^+(1)$ |
| 47, 48 | 2 | 112.95 | $E_{2u}, \Gamma_6^-(1)$ |
| 49, 50 | 2 | 113.14 | $E_{1g}, \Gamma_5^+(1)$ |
| 51, 52 | 2 | 113.98 | $E_{1u}, \Gamma_5^-(1)$ |
| 53 | 1 | 118.83 | $B_{2g}, \Gamma_4^+(1)$ |
| 54, 55 | 2 | 122.79 | $E_{2g}, \Gamma_6^+(1)$ |
| 56, 57 | 2 | 122.90 | $E_{1u}, \Gamma_5^-(1)$ |
| 58 | 1 | 123.83 | $A_{1g}, \Gamma_1^+(1)$ |
| 59, 60 | 2 | 126.54 | $E_{2u}, \Gamma_6^-(1)$ |
| 61 | 1 | 128.18 | $A_{2u}, \Gamma_2^-(1)$ |
| 62, 63 | 2 | 128.85 | $E_{2g}, \Gamma_6^+(1)$ |
| 64 | 1 | 129.50 | $B_{1u}, \Gamma_3^-(1)$ |
| 65 | 1 | 135.75 | $B_{2u}, \Gamma_4^-(1)$ |
| 66 | 1 | 136.69 | $A_{1g}, \Gamma_1^+(1)$ |
| 67, 68 | 2 | 137.84 | $E_{1u}, \Gamma_5^-(1)$ |
| 69 | 1 | 144.90 | $B_{2g}, \Gamma_4^+(1)$ |
| 70, 71 | 2 | 165.93 | $E_{1u}, \Gamma_5^-(1)$ |
| 72 | 1 | 169.21 | $A_{2g}, \Gamma_2^+(1)$ |
| 73, 74 | 2 | 172.19 | $E_{2g}, \Gamma_6^+(1)$ |
| 75 | 1 | 172.93 | $B_{1u}, \Gamma_3^-(1)$ |
| 76, 77 | 2 | 174.56 | $E_{1u}, \Gamma_5^-(1)$ |
| 78 | 1 | 187.69 | $A_{1g}, \Gamma_1^+(1)$ |
| 79, 80 | 2 | 199.18 | $E_{2g}, \Gamma_6^+(1)$ |
| 81 | 1 | 203.43 | $B_{1u}, \Gamma_3^-(1)$ |
| 82, 83 | 2 | 212.30 | $E_{1u}, \Gamma_5^-(1)$ |
| 84, 85 | 2 | 220.63 | $E_{2g}, \Gamma_6^+(1)$ |
| 86, 87 | 2 | 232.00 | $E_{2g}, \Gamma_6^+(1)$ |
| 88 | 1 | 235.78 | $A_{1g}, \Gamma_1^+(1)$ |
| 89 | 1 | 240.72 | $B_{1u}, \Gamma_3^-(1)$ |
| 90 | 1 | 241.02 | $A_{2g}, \Gamma_2^+(1)$ |
| 91 | 1 | 246.49 | $A_{2u}, \Gamma_2^-(1)$ |
| 92 | 1 | 252.53 | $B_{2g}, \Gamma_4^+(1)$ |
| 93, 94 | 2 | 254.71 | $E_{1u}, \Gamma_5^-(1)$ |
| 95 | 1 | 260.43 | $B_{2u}, \Gamma_4^-(1)$ |
| 96, 97 | 2 | 262.14 | $E_{1g}, \Gamma_5^+(1)$ |
| 98 | 1 | 262.60 | $A_{2u}, \Gamma_2^-(1)$ |
| 99, 100 | 2 | 268.59 | $E_{2u}, \Gamma_6^-(1)$ |
| 101, 102 | 2 | 274.83 | $E_{1u}, \Gamma_5^-(1)$ |
| 103, 104 | 2 | 276.24 | $E_{1g}, \Gamma_5^+(1)$ |
| 105 | 1 | 296.24 | $B_{1g}, \Gamma_3^+(1)$ |
| 106 | 1 | 298.20 | $B_{2u}, \Gamma_4^-(1)$ |
| 107, 108 | 2 | 306.00 | $E_{2u}, \Gamma_6^-(1)$ |

**Note S2: Excited phonon modes and laser-induced lattice distortions**

To elucidate the lattice dynamics following photoexcitation, we analyze the time evolution of each phonon branch by projecting the ionic displacements along the MD trajectory onto the phonon eigenvectors computed from the ISD phase [Fig. S2(b)]. The projection intensity is defined as $I_i(t) = \sum_j \boldsymbol{e}_i \cdot \boldsymbol{\mu}_j(t)$, where $\boldsymbol{e}_i$ is the $i$-th phonon eigenvector, $\boldsymbol{\mu}_j$ is the displacement vector of atom $j$ at time $t$ relative to its equilibrium position at $t = 0$. As shown in Fig. S3(a), in addition to the dominant out-of-plane oscillation of Sb2 atoms (mode index $n = 66$, ~136.7 cm$^{-1}$), the laser induced lattice distortion also involves substantial in-plane displacements of V atoms and their neighboring Sb2 atoms. These collective motions are primarily composed of phonons with either $A_{1g}$ or $E_{2g}$ symmetry, as confirmed by group-theoretical analysis (Table S1). The corresponding real-space patterns exhibit breathing-type and chiral oscillations [Fig. S3(b-d)], and their frequencies match well with experimental measurement [Table S2]. Within the $D_{6h}$ point group, the triple-$\mathbf{Q}_M$ mode decomposes into $3M_1^+ \rightarrow A_{1g} \oplus E_{2g}$, which justifies our further projection of ionic displacements onto the phonon eigenmodes associated with each single-$\mathbf{Q}_M$ distortion.

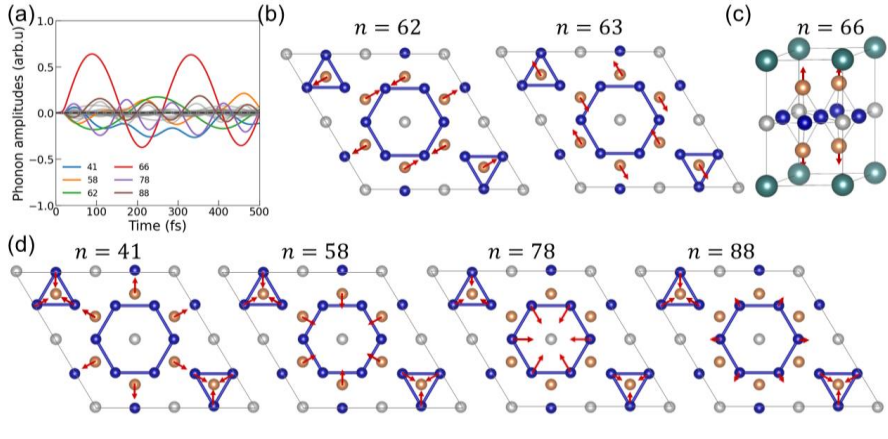

FIG. S3. Time-dependent phonon excitation under a peak electric field $E_0$=0.04 V/Å. Six dominant phonon modes are highlighted, while the remaining 102 modes are shown as gray traces. (b) Real-space vibration pattern of the doubly degenerate $E_{2g}$ modes (mode indices $n = 62, 63$), featuring counter-rotating circular motions of Sb2 atoms within the $ab$-plane. (c-d) Vibration patterns of the $A_{1g}$ modes: (c) out-of-plane oscillations of Sb2 atoms and (d) in-plane breathing-type motions of V and Sb2 atoms.

**TABLE. S2.** Frequency (in cm$^{-1}$) of the Raman-active modes in the 2 × 2 × 1 ISD state of RbV$_3$Sb$_5$ and CsV$_3$Sb$_5$ [16].

|           | n=41  | n=58  | n=66  | n=78  | n=88  |
|-----------|-------|-------|-------|-------|-------|
| This work | 96.5  | 123.8 | 136.7 | 187.7 | 235.8 |
| Exp.[16]. | 104.1 | 127.2 | 136.7 | 197.4 | 241.5 |

**Note S3: Dependence of lattice distortion on laser parameters**

To elucidate the polarization dependence of ultrafast lattice distortion, we analyze the temporal evolution of selected V-V bond lengths in response to linearly polarized laser excitation. Specifically, we focus on two representative V atoms located at the vertices of hexagons and triangles, respectively, and track their nearest-neighbor bond lengths [Fig. S4]. The most pronounced bond modulations are observed along the direction of laser polarization. For instance, when the electric field is polarized along the crystallographic $a$-axis (denoted as $E_1$), the bond lengths of V7-V4 and V7-V8 exhibit greater deviations compared to V7-V3 and V7-V9. This directional sensitivity leads to an alternating pattern in bonding and anti-bonding interactions, governed by equilibrium V-V distance of 2.72 Å in the pristine phase.

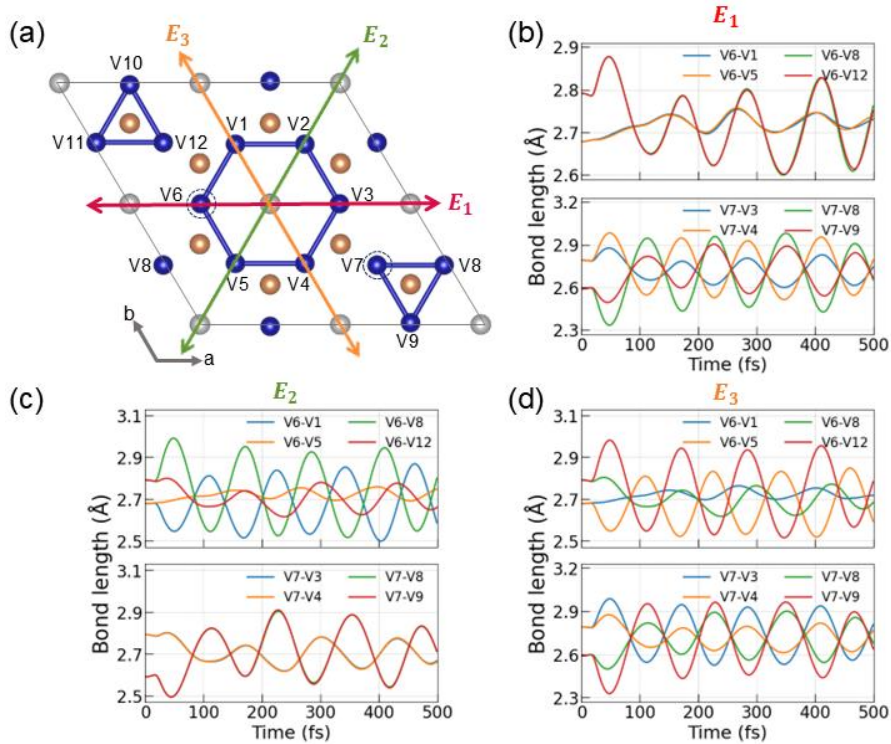

FIG. S4. (a) Schematic illustration of the ISD phase under irradiation by linearly polarized laser pulses with $E_0$=0.04 V/Å. (b-d) Temporal evolution of selective V-V bond lengths under different laser polarization orientations, showing pronounced anisotropic lattice response aligned with the electric field direction.

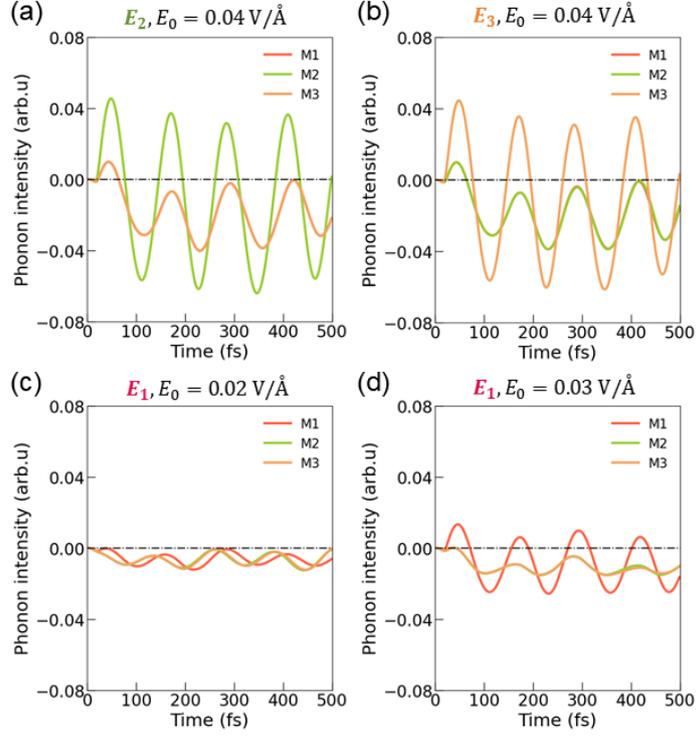

FIG. S5. Time-resolved atomic displacements projected onto three single-$\mathbf{Q}_M$ modes under various laser parameters. The results highlight the polarization- and intensity-dependent excitation selectivity of distinct phonon branches.

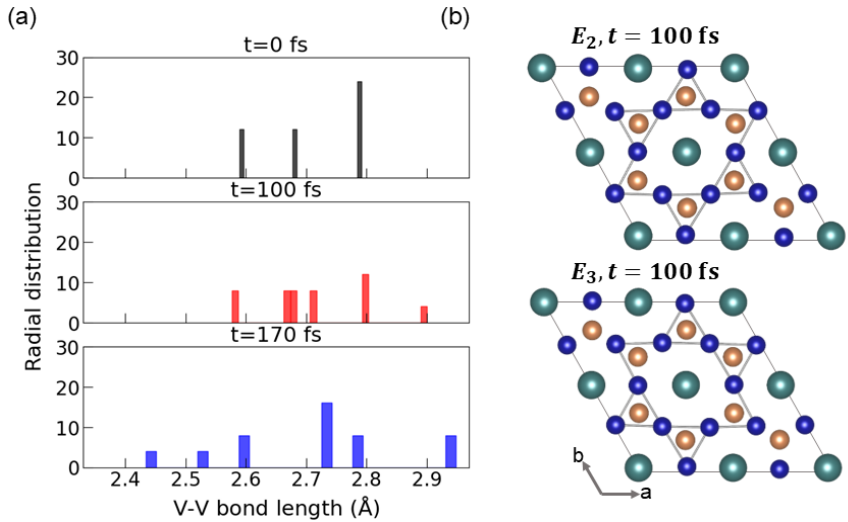

FIG. S6. (a) Comparison of the V-V bond length distributions between the ISD and TSB states. Despite the anisotropic excitation, the resulting TSB states exhibit nearly identical distortion profiles across different polarizations. (b-c) Snapshots of the TSB states at 100 fs under $E_0$=0.04 V/Å with laser polarization along the $E_2$ and $E_3$ directions, respectively.

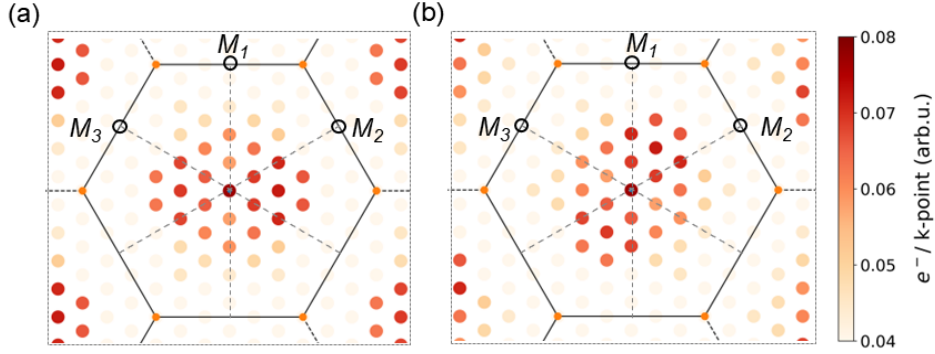

FIG. S7. Momentum-resolved distribution of photoexcited electrons after the laser pulse, shown as false-color maps to illustrate the direction-selective coupling of light. (a) Polarization along the $E_1$ direction; (b) Polarization along the $E_3$ direction.

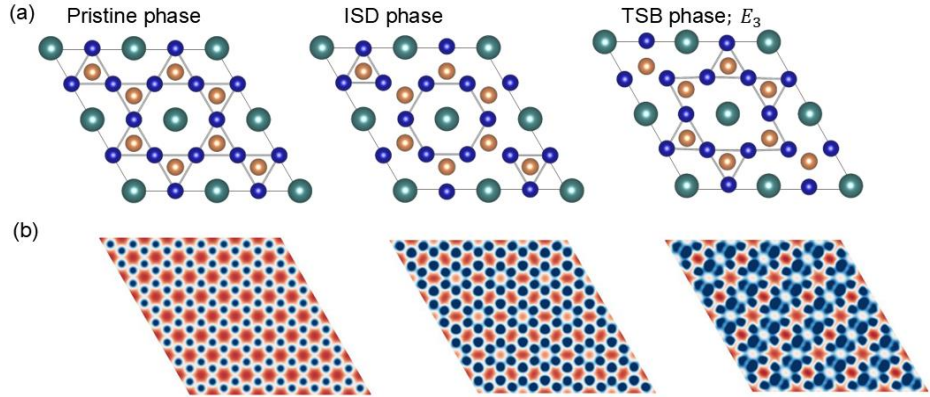

FIG. S8. (a) Atomic configurations of RbV$_3$Sb$_5$ in the pristine state, ISD state and TSB state at 100 fs following laser excitation ($E_0$=0.04 V/Å, polarization along the $E_3$ direction). (b) Simulated STM images corresponding to the three structural phases, reveal distinct signatures.

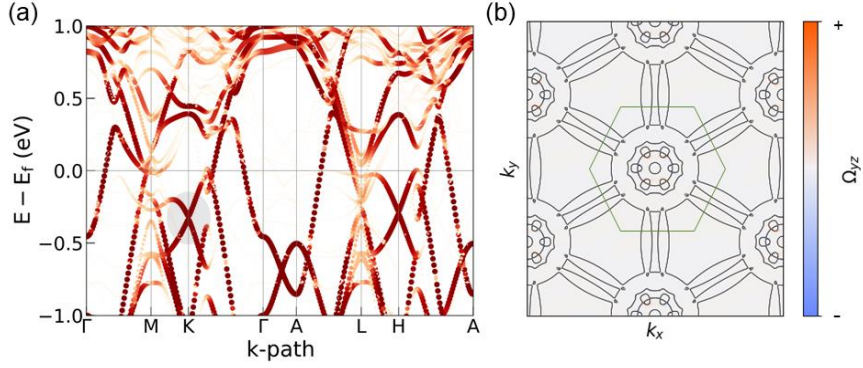

FIG. S9. (a) Electronic band structure of the ISD state. (b) Fermi surface (black lines) and Berry curvature distribution across the Brillouin zone, with the green hexagon outlines the first Brillouin zone.

**Note S4. The tight-binding model**

To confirm the origin of the bandgap opening and magnetic excitation, we construct an effective tight-binding (TB) model for the space group $P6/mmm$ (BNS No. 191.234) using the MagneticTB package. We choose the $d_{z^2}$ orbital at the 3f Wyckoff position and consider nearest-neighbor hopping within a $2 \times 2 \times 1$ Kagome vanadium layer. Owing to the uniform V-V bond lengths and symmetric interactions, the resulting $12 \times 12$ Hamiltonian for the pristine phase can be written as

$$H_0 = \sum_i e_i c_i^\dagger c_i + \sum_{\langle i,j \rangle} t c_i^\dagger c_j ,$$

where $t$ denotes the interaction hopping amplitude along the three edges of each corner-sharing triangle. The explicit matrix form of the hooping term is given by

$$\begin{pmatrix}
0 & e^{\frac{iky}{4}}t & 0 & 0 & 0 & e^{i\left(-\frac{kx}{4}-\frac{ky}{4}\right)}t & 0 & 0 & 0 & 0 & e^{-\frac{iky}{4}}t & e^{i\left(\frac{kx}{4}+\frac{ky}{4}\right)}t \\
e^{-\frac{iky}{4}}t & 0 & e^{\frac{ikx}{4}}t & 0 & 0 & 0 & e^{\frac{iky}{4}}t & 0 & 0 & 0 & 0 & e^{\frac{ikx}{4}}t \\
0 & e^{\frac{ikx}{4}}t & 0 & e^{i\left(-\frac{kx}{4}-\frac{ky}{4}\right)}t & 0 & 0 & e^{i\left(\frac{kx}{4}+\frac{ky}{4}\right)}t & e^{-\frac{ikx}{4}}t & 0 & 0 & 0 & 0 \\
0 & 0 & e^{i\left(\frac{kx}{4}+\frac{ky}{4}\right)}t & 0 & e^{\frac{iky}{4}}t & 0 & 0 & e^{\frac{iky}{4}}t & e^{i\left(-\frac{kx}{4}-\frac{ky}{4}\right)}t & 0 & 0 & 0 \\
0 & 0 & 0 & e^{\frac{iky}{4}}t & 0 & e^{\frac{ikx}{4}}t & 0 & 0 & e^{-\frac{ikx}{4}}t & e^{-\frac{iky}{4}}t & 0 & 0 \\
e^{i\left(\frac{kx}{4}+\frac{ky}{4}\right)}t & 0 & 0 & 0 & e^{\frac{ikx}{4}}t & 0 & 0 & 0 & 0 & e^{i\left(-\frac{kx}{4}-\frac{ky}{4}\right)}t & e^{\frac{ikx}{4}}t & 0 \\
0 & e^{-\frac{iky}{4}}t & e^{i\left(-\frac{kx}{4}-\frac{ky}{4}\right)}t & 0 & 0 & 0 & 0 & 0 & e^{i\left(\frac{kx}{4}+\frac{ky}{4}\right)}t & 0 & e^{\frac{iky}{4}}t & 0 \\
0 & 0 & e^{\frac{ikx}{4}}t & e^{\frac{iky}{4}}t & 0 & 0 & 0 & 0 & 0 & e^{\frac{iky}{4}}t & 0 & e^{-\frac{ikx}{4}}t \\
0 & 0 & 0 & e^{i\left(\frac{kx}{4}+\frac{ky}{4}\right)}t & e^{\frac{ikx}{4}}t & 0 & e^{i\left(-\frac{kx}{4}-\frac{ky}{4}\right)}t & 0 & 0 & 0 & e^{-\frac{ikx}{4}}t & 0 \\
0 & 0 & 0 & 0 & e^{\frac{iky}{4}}t & e^{i\left(\frac{kx}{4}+\frac{ky}{4}\right)}t & 0 & e^{-\frac{iky}{4}}t & 0 & 0 & 0 & e^{i\left(-\frac{kx}{4}-\frac{ky}{4}\right)}t \\
e^{\frac{iky}{4}}t & 0 & 0 & 0 & 0 & e^{-\frac{ikx}{4}}t & e^{-\frac{iky}{4}}t & 0 & e^{\frac{ikx}{4}}t & 0 & 0 & 0 \\
e^{i\left(-\frac{kx}{4}-\frac{ky}{4}\right)}t & e^{-\frac{ikx}{4}}t & 0 & 0 & 0 & 0 & 0 & e^{\frac{ikx}{4}}t & 0 & e^{i\left(\frac{kx}{4}+\frac{ky}{4}\right)}t & 0 & 0
\end{pmatrix},$$

Upon photoexcitation, the Hamiltonian evolves as $\Delta H = H_1 + H_2$, where $H_1$ and $H_2$ describe corrections arising from the lattice distortion that forms the ISD state and from the excitation of single-$Q_M$ mode, respectively. These terms are expressed as

$$H_1 =$$

$$\begin{pmatrix}
0 & e^{\frac{iky}{4}}\Delta t_1 & 0 & 0 & 0 & e^{i\left(-\frac{kx}{4}-\frac{ky}{4}\right)}\Delta t_1 & 0 & 0 & 0 & 0 & 0 & 0 \\
e^{-\frac{iky}{4}}\Delta t_1 & 0 & e^{\frac{ikx}{4}}\Delta t_1 & 0 & 0 & 0 & 0 & 0 & 0 & 0 & 0 & 0 \\
0 & e^{-\frac{ikx}{4}}\Delta t_1 & 0 & e^{i\left(-\frac{kx}{4}-\frac{ky}{4}\right)}\Delta t_1 & 0 & 0 & 0 & 0 & 0 & 0 & 0 & 0 \\
0 & 0 & e^{i\left(\frac{kx}{4}+\frac{ky}{4}\right)}\Delta t_1 & 0 & e^{-\frac{iky}{4}}\Delta t_1 & 0 & 0 & 0 & 0 & 0 & 0 & 0 \\
0 & 0 & 0 & e^{\frac{iky}{4}}\Delta t_1 & 0 & e^{\frac{ikx}{4}}\Delta t_1 & 0 & 0 & 0 & 0 & 0 & 0 \\
e^{i\left(\frac{kx}{4}+\frac{ky}{4}\right)}\Delta t_1 & 0 & 0 & 0 & e^{-\frac{ikx}{4}}\Delta t_1 & 0 & 0 & 0 & 0 & 0 & 0 & 0 \\
0 & 0 & 0 & 0 & 0 & 0 & 0 & e^{i\left(\frac{kx}{4}+\frac{ky}{4}\right)}\Delta t_1 & 0 & e^{\frac{iky}{4}}\Delta t_1 & 0 \\
0 & 0 & 0 & 0 & 0 & 0 & 0 & 0 & e^{\frac{iky}{4}}\Delta t_1 & 0 & e^{-\frac{ikx}{4}}\Delta t_1 \\
0 & 0 & 0 & 0 & 0 & 0 & e^{i\left(-\frac{kx}{4}-\frac{ky}{4}\right)}\Delta t_1 & 0 & 0 & 0 & e^{-\frac{ikx}{4}}\Delta t_1 & 0 \\
0 & 0 & 0 & 0 & 0 & 0 & 0 & e^{-\frac{iky}{4}}\Delta t_1 & 0 & 0 & 0 & e^{i\left(-\frac{kx}{4}-\frac{ky}{4}\right)}\Delta t_1 \\
0 & 0 & 0 & 0 & 0 & 0 & e^{-\frac{iky}{4}}\Delta t_1 & 0 & e^{\frac{ikx}{4}}\Delta t_1 & 0 & 0 & 0 \\
0 & 0 & 0 & 0 & 0 & 0 & 0 & e^{\frac{ikx}{4}}\Delta t_1 & 0 & e^{i\left(\frac{kx}{4}+\frac{ky}{4}\right)}\Delta t_1 & 0 & 0
\end{pmatrix},$$

$$H_2 =$$

$$\begin{pmatrix}
0 & 0 & 0 & 0 & 0 & 0 & 0 & 0 & 0 & 0 & e^{-\frac{iky}{4}}\Delta t_2 & 0 \\
0 & 0 & 0 & 0 & 0 & 0 & e^{\frac{iky}{4}}\Delta t_2 & 0 & 0 & 0 & 0 & 0 \\
0 & 0 & 0 & 0 & 0 & 0 & 0 & 0 & 0 & 0 & 0 & 0 \\
0 & 0 & 0 & 0 & 0 & 0 & 0 & e^{\frac{iky}{4}}\Delta t_2 & 0 & 0 & 0 & 0 \\
0 & 0 & 0 & 0 & 0 & 0 & 0 & 0 & 0 & e^{-\frac{iky}{4}}\Delta t_2 & 0 & 0 \\
0 & 0 & 0 & 0 & 0 & 0 & 0 & 0 & 0 & 0 & 0 & 0 \\
0 & e^{-\frac{iky}{4}}\Delta t_2 & 0 & 0 & 0 & 0 & 0 & 0 & 0 & 0 & 0 & 0 \\
0 & 0 & 0 & e^{-\frac{iky}{4}}\Delta t_2 & 0 & 0 & 0 & 0 & 0 & 0 & 0 & 0 \\
0 & 0 & 0 & 0 & 0 & 0 & 0 & 0 & 0 & 0 & 0 & 0 \\
0 & 0 & 0 & 0 & e^{\frac{iky}{4}}\Delta t_2 & 0 & 0 & 0 & 0 & 0 & 0 & 0 \\
e^{\frac{iky}{4}}\Delta t_2 & 0 & 0 & 0 & 0 & 0 & 0 & 0 & 0 & 0 & 0 & 0 \\
0 & 0 & 0 & 0 & 0 & 0 & 0 & 0 & 0 & 0 & 0 & 0
\end{pmatrix},$$

with $\Delta t_1$ and $\Delta t_2$ controlling the respective amplitudes of the distortions. A finite $\Delta t_1$ drives transition from the pristine phase to the ISD phase, while $H_2$ introduces anisotropic hopping with magnitude $\Delta t_2$. For example, the V atoms labeled 1, 6 and 11 form a corner-sharing triangle [Fig. S10(a)]. When the $M_3$ mode is excited, the hopping term between atoms 1 and 11 is modified. Figure S10 (d-f) shows the resulting band structures computed with fixed $t = 1$ while varying $\Delta t_1$ and $\Delta t_2$ to emulate the structural evolution observed in dynamical simulations.

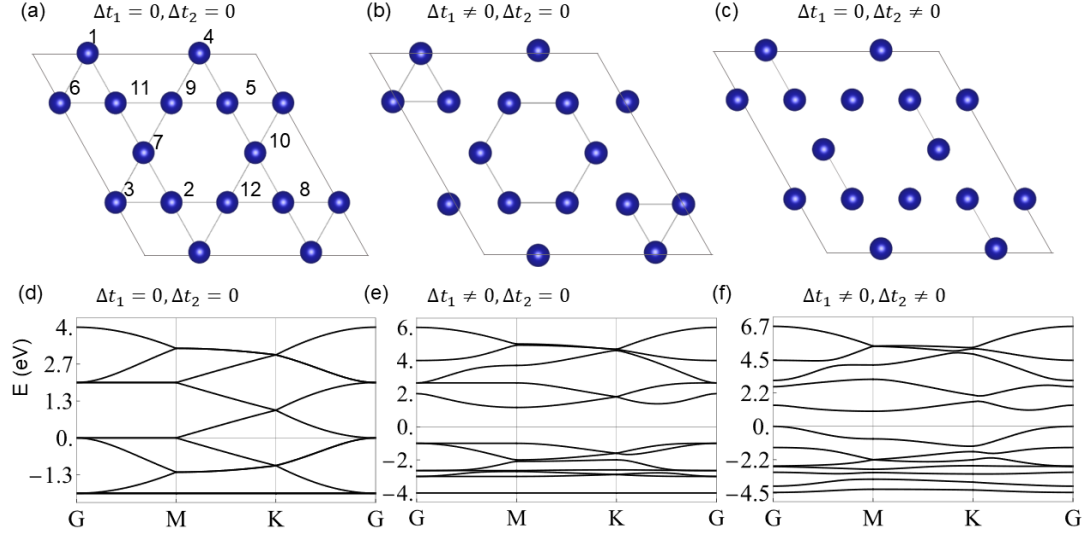

FIG. S10. (a-b) Atomic structures of the TB model with different parameters: (a) Kagome vanadium plane in the pristine phase. (b) Kagome vanadium plane in the ISD phase. (c) Kagome vanadium plane with an excited $M_3$ mode. The numbers in (a) indicate atomic indices. (d-f) Band structures from the corresponding TB models: (d) Pristine phase; (e) ISD phase; (f) ISD phase with the additional excitation of the $M_3$ mode.

**Note S5. Magnetic excitation in the pristine phase**

To assess the magnetic response of the pristine $AV_3Sb_5$ structure, we investigate two distinct types of lattice perturbation: dynamics atomic displacements induced by excitation of the $M_1$ mode and static uniaxial strain applied along the $a$-axis. The spin moment distributions under these conditions are displayed in Fig. S11. Figure S11 (a-b) reveal the emergence of a robust ferrimagnetic order, similar to that of the TSB state [Fig. 3(b) in the main text], upon application of both positive and negative atomic displacements along the $M_1$ eigenvector defined in Fig. 1(b). The amplitude of displacement for each V atoms is ~0.2 Å, corresponding to ~2% of the in-plane lattice constant, while the cell dimensions remain fixed. This distortion yields a net magnetic moment of 1.44 $\mu_B$/cell. In contrast, applying a ±2% uniaxial strain while holding atomic positions fixed results in only a negligible spin polarization (~0.02 $\mu_B$/cell), highlighting that V-V bond dimerization, rather than macroscopic strain, or anisotropic alone, is the primary driver of the spin polarization induced by laser excitation.

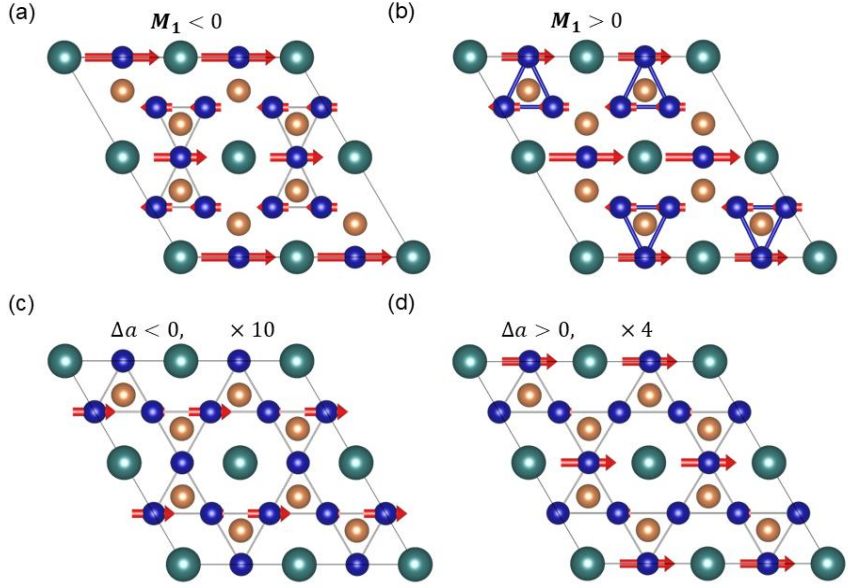

FIG. S11. Real-space spin moment distribution on V atoms in the pristine phase under different lattice distortion scenarios. (a-b) Ferrimagnetic order induced by dynamic displacements along the $M_1$ phonon mode direction, with fixed lattice constants. (c-d) Nearly vanishing spin polarization under $\pm 2\%$ uniaxial strain along $a$-axis with fixed atomic position. Spin vectors are exaggerated for clarity in both panels.

**Note S6. Magnetic excitation in the three-dimensional CDW phases**

The stacking degree of freedom plays a crucial role in modulating electronic and magnetic properties of $A$V$_3$Sb$_5$, including superconductivity, charge bond order and electronic nematicity. A π-phase shift between adjacent layers has been proposed, wherein the CDW distortion pattern is laterally shifted by half a unit cell in-plane between neighboring layers. Raman measurements suggest the presence of three energetically equivalent domains, each oriented ~120° apart[17], as shown in Fig. S12 (a-b). Based on this, we model a uniform laser excitation scenario in which the ISD phase forms identically in each layer, consistent with the limited optical penetration depth and surface-sensitive experimental conditions. Stacking the TSB phase [from Fig. 2(b)] using three possible relative arrangement confirms that the ferrimagnetic spin configuration remains robust in all cases.

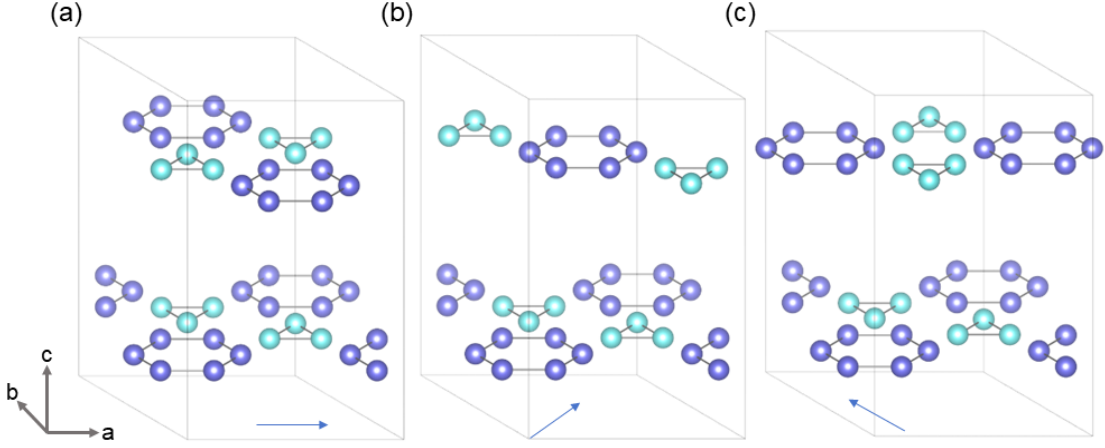

FIG. S12. Schematic representations of three-dimensional CDW stacking with π-phase shift. Each upper layer is laterally shifted by half a lattice constant in three directions (indicated by blue arrows) relative to the bottom layer, forming distinct stacking configurations.

**Note S7. Potential for realizing artificially designed structural chirality in $A$V$_3$Sb$_5$**

The emergence of chiral charge order and superconductively in $A$V$_3$Sb$_5$ are typically attributed to electronically driven symmetry breaking, rather than a chiral atomic lattice[18]. However, helicity-dependent photocurrent measurements have revealed an intrinsic chiral response[19]. Given the potential coexistence of multiple charge-ordering patterns, the emergence of a chiral crystal may arise from their collective superposition. While the microscopic origin of this intrinsically chirality and the precise interlayer stacking sequence remain elusive, the demonstrated tunability of symmetry breaking at the monolayer level underscores the feasibility of artificially designing chiral lattice architecture through optical control.

In the ISD phase, several crystalline symmetries, including $\mathcal{C}_6$, $\mathcal{C}_3$, $\mathcal{C}_2$, $\mathcal{M}_{100}$, $\mathcal{M}_{010}$, $\mathcal{M}_{001}$, $\mathcal{M}_{210}$, $\mathcal{M}_{120}$, $\mathcal{M}_{110}$, $\mathcal{M}_{1\bar{1}0}$ and inversion are preserved. Upon selective enhancement of a single-$\mathbf{Q}_M$ modes, *e.g.*, $M_1$, the symmetry lowers to the group *Cmmm*, retaining only $\mathcal{C}_2$, $\mathcal{M}_{010}$, $\mathcal{M}_{100}$, $\mathcal{M}_{001}$ and inversion. If different single-$\mathbf{Q}_M$ modulations are imposed across adjacent layers, all mirror and inversion symmetries are broken, yielding a chiral three-dimensional structure. We construct a model comprising three monolayers stacked in an $M_1$-$M_2$-$M_3$ sequence, corresponding to the chiral space group $P6_222$. This chiral configuration could potentially be realized by irradiating the materials with circularly polarized light along the stacking axis. Due to the finite penetration depth, the local polarization orientation evolves across the sample depth, favoring a specific stacking chirality [20].

The circular photogalvanic effect (CPGE), which probes the difference in photocurrent between right- (RCP) and left-circularly polarized (LCP) light, serves as a sensitive diagnostic of structural chirality [19]. As shown in Fig. S13, CPGE along the $z$-axis exhibits strong circular dichroism, with RCP and LCP generating out-of-plane photocurrent of equal magnitude but opposite sign. Such designable structures offer a promising platform for systematically investigating the interplay between structural chirality and correlation-driven electronic orders.

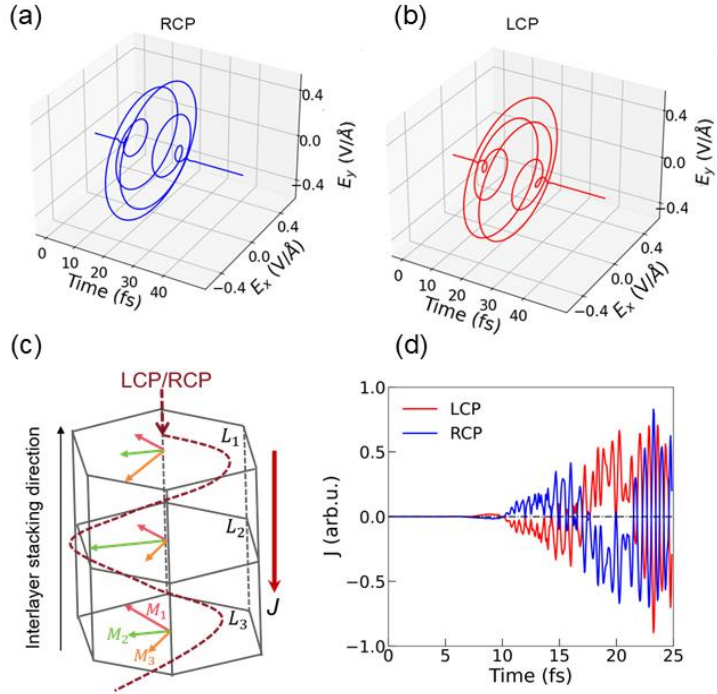

FIG. S13. (a-b) Electric field waveforms of right- and left-circularly polarized (RCP and LCP) laser pulses. (c) Schematic of laser-driven formation of chiral stacking, enabled by depth-dependent electric field orientation. (d) Time-resolved circular photogalvanic current along the interlayer direction, showing opposite responses to RCP and LCP illumination.

[1] G. Kresse and D. Joubert, *From ultrasoft pseudopotentials to the projector augmented-wave method,* Phys. Rev. B **59**, 1758 (1999).

[2] G. Kresse and J. Furthmüller, *Efficient iterative schemes for ab initio total-energy calculations using a plane-wave basis set,* Phys. Rev. B **54**, 11169 (1996).

[3] G. Kresse and J. Furthmüller, *Efficiency of ab-initio total energy calculations for metals and semiconductors using a plane-wave basis set,* Comp. Mater. Sci. **6**, 15 (1996).

[4] J. P. Perdew, K. Burke, and M. Ernzerhof, *Generalized gradient approximation made simple,* Phys. Rev. Lett. **77**, 3865 (1996).

[5] S. Grimme, J. Antony, S. Ehrlich, and H. Krieg, *A consistent and accurate ab initio parametrization of density functional dispersion correction (DFT-D) for the 94 elements H-Pu,* J. Chem. Phys. **132**, 154104 (2010).

[6] V. Wang, N. Xu, J.-C. Liu, G. Tang, and W.-T. Geng, *VASPKIT: A user-friendly interface facilitating high-throughput computing and analysis using VASP code,* Computer Physics Communications **267**, 108033 (2021).

[7] Z. Zhang, Z.-M. Yu, G.-B. Liu, and Y. Yao, *MagneticTB: A package for tight-binding model of magnetic and non-magnetic materials,* Computer Physics Communications **270** (2022).

[8] A. Togo and I. Tanaka, *First principles phonon calculations in materials science,* Scripta Materialia **108**, 1 (2015).

[9] Z.-M. Y. Zeying Zhang, Gui-Bin Liu, and Yugui Yao, *A phonon irreducible representations calculator,* arXiv preprint arXiv:2201.11350 (2022).

[10] K. Choudhary, K. F. Garrity, C. Camp, S. V. Kalinin, R. Vasudevan, M. Ziatdinov, and F. Tavazza, *Computational scanning tunneling microscope image database,* Sci Data **8**, 57 (2021).

[11] Q. Wu, S. Zhang, H.-F. Song, M. Troyer, and A. A. Soluyanov, *WannierTools: An open-source software package for novel topological materials,* Computer Physics Communications **224**, 405 (2018).

[12] A. A. Mostofi, J. R. Yates, Y.-S. Lee, I. Souza, D. Vanderbilt, and N. Marzari, *wannier90: A tool for obtaining maximally-localised Wannier functions,* Computer physics communications **178**, 685 (2008).

[13] M. Guan, D. Chen, S. Hu, H. Zhao, P. You, and S. Meng, *Theoretical Insights into Ultrafast Dynamics in Quantum Materials,* Ultrafast Science **2022**, 9767251 (2022).

[14] C. Lian, M. Guan, S. Hu, J. Zhang, and S. Meng, *Photoexcitation in Solids: First-Principles Quantum Simulations by Real-Time TDDFT,* Adv. Theory. Simul. **1**, 1800055 (2018).

[15] Y. Xing *et al.*, *Optical manipulation of the charge-density-wave state in RbV3Sb5,* Nature **631**, 60 (2024).

[16] G. Liu *et al.*, *Observation of anomalous amplitude modes in the kagome metal CsV3Sb5,* Nat. Commun. **13**, 3461 (2022).

[17] F. Jin, W. Ren, M. Tan, M. Xie, B. Lu, Z. Zhang, J. Ji, and Q. Zhang, *pi Phase Interlayer Shift and Stacking Fault in the Kagome Superconductor CsV3Sb5,* Phys. Rev. Lett. **132**, 066501 (2024).

[18] S. Y. Xu *et al.*, *Spontaneous gyrotropic electronic order in a transition-metal dichalcogenide,* Nature **578**, 545 (2020).

[19] Z. J. Cheng *et al.*, *Broken symmetries associated with a Kagome chiral charge order,* Nat. Commun. **16**, 3782 (2025).

[20] J. Ishioka, Y. H. Liu, K. Shimatake, T. Kurosawa, K. Ichimura, Y. Toda, M. Oda, and S. Tanda, *Chiral charge-density waves,* Phys. Rev. Lett. **105**, 176401 (2010).